\documentclass[sigconf, nonacm]{acmart}

\usepackage{tikz}
\usepackage{amsmath}
\usepackage{filecontents}

\usepackage[linesnumbered,boxed,ruled,commentsnumbered]{algorithm2e}
\usepackage{amsmath}
\usepackage{comment}
\usepackage{amssymb}
\usepackage[most]{tcolorbox}
\usepackage{graphicx}
\usepackage{subfigure}
\usepackage{algorithmic}
\usepackage{xcolor}
\newcommand{\cmt}[1]{\textcolor[rgb]{0,0.5,0}{// #1}} 
\usepackage{url}
\usepackage{xspace}
\usepackage{balance}
\usepackage{colortbl} 
\usepackage{booktabs}
\usepackage{tabularx} 
\usepackage{array} 

\usepackage{amsthm}

\usepackage{enumitem}
\usepackage{arydshln}
\usepackage{caption}
\setlist[itemize]{leftmargin=*}
\usepackage{booktabs}
\usepackage{siunitx}
\definecolor{darkgreen}{rgb}{0.0, 0.5, 0.0}

\newtheorem{example}{Example}

\newtheorem{definition}{Definition}

\newtheorem{remark}{Remark}

\hypersetup{
  colorlinks,
  linkcolor={blue!70!green},
  citecolor={green!70!blue},
  urlcolor={orange!70!red}
}
\newcommand{\mypara}[1]{\smallskip\noindent\textbf{#1.}\xspace}
\newcommand{\method}{\textsf{FedSGT}\xspace}

\newtcolorbox{takeawaybox}{
  enhanced,
  colback=gray!7,
  colframe=white,                
  left=2pt,right=2pt,top=1pt,bottom=1pt, 
  boxsep=0pt,                   
  before skip=2pt,after skip=2pt,
  borderline west={1.2pt}{0pt}{black!40!blue}, 
  sharp corners                 
}

\begin{document}
\title{\method: Exact Federated Unlearning via Sequential Group-based Training}

\author{Bokang Zhang}
\affiliation{%
  \institution{Emory University}
}
\email{bokang.zhang@emory.edu}

\author{Hong Guan}
\affiliation{%
  \institution{Arizona State University}
}
\email{hguan6@asu.edu}

\author{Hong Kyu Lee}
\affiliation{%
  \institution{Emory University}
}
\email{hong.kyu.lee@emory.edu}

\author{Ruixuan Liu}
\affiliation{%
  \institution{Emory University}
}
\email{ruixuan.liu2@emory.edu}

\author{Jia Zou}
\affiliation{%
  \institution{Arizona State University}
}
\email{Jia.Zou@asu.edu}

\author{Li Xiong}
\affiliation{%
  \institution{Emory University}
}
\email{lxiong@emory.edu}


\begin{abstract}
Federated Learning (FL) enables collaborative, privacy-preserving model training, but supporting the ``Right to be Forgotten" is especially challenging because data influences the model through distributed and interleaved client updates.
Existing exact unlearning methods 
typically require frequent retraining from scratch, resulting in high communication cost and long service downtime.
To address this, we propose \underline{Fed}erated \underline{S}equential \underline{G}roup-based \underline{T}raining (\method), an exact unlearning framework for FL. 
\method partitions the data into uniform groups, and each client may participate in multiple groups. To control communication overhead, each client can limit the number of groups it contributes to. 
\method then trains multiple sequences of Parameter-Efficient Fine-Tuning (PEFT) modules, each corresponding to a different group permutation.
Since the PEFT modules are lightweight and maintained server-side, \method isolates the influence of different data groups into independent modules without incurring significant storage overhead and communication cost. 
Exact unlearning is thus achieved instantly by deactivating the modules corresponding to the group containing the unlearned data. 
Furthermore, using multiple training sequences helps maintain high model utility as deletion requests accumulate.
We provide a rigorous theoretical analysis of both the deletion rate—expected number of deletions before retraining is needed—and the expected model performance.  
Experiments on various tasks demonstrate that \method achieves 
a significantly longer service maintenance under multiple unlearning requests while maintaining comparable learning performance and training efficiency to other exact unlearning baselines. 
Extensive ablation studies validate the robustness of our method across a wide range of parameter settings.
\end{abstract}

\maketitle

\section{Introduction}

Federated Learning (FL)~\cite{mcmahan2017communication, li2020federated, zhang2021survey, kairouz2021advances} has emerged as a dominant privacy-preserving machine learning paradigm, enabling multiple data silos to collaboratively train a global model without sharing their local raw data. 
This distributed nature has led to its widespread adoption in privacy-sensitive domains such as finance and healthcare.
Recently, the increasing stringency of data privacy regulations, such as the GDPR~\cite{voigt2017eu}, has granted users with the \textit{Right to be Forgotten}, mandating the removal of their data contributions from models. 
Merely deleting data from a database is insufficient, as model parameters may still retain ``memories'' of the deleted data. 
Consequently, Machine Unlearning (MU)~\cite{nguyen2025survey, kurmanji2024machine, vatter2023evolution, xia2023equitable} techniques have been developed to efficiently erase the influence of specific data from a trained model. 
Applying this requirement to FL creates the challenging problem of Federated Unlearning (FU), which aims to remove specific clients' data from the global model as if it had never been trained on the data.

Existing MU approaches are broadly categorized into approximate unlearning~\cite{gupta2021adaptive,lin2023erm, zhang2024contrastive}, and exact unlearning~\cite{bourtoule2021machine, chen2022recommendation, yan2022arcane, xia2025edge}.
Approximate methods modify or fine-tune the model to weaken the impact of the forgotten data, but they typically fail to provide a theoretical guarantee of complete removal, which is unsuitable for applications demanding stringent security and privacy.
In contrast, exact unlearning ensures that the unlearned model behaves indistinguishable from one retrained from scratch on the remaining data, providing strong verifiability guarantees.

In the federated setting, achieving exact unlearning remains especially challenging because data influences the model through distributed and interleaved client updates.
Most existing FU approaches adopt approximate strategies—such as gradient ascent~\cite{pan2025federated} or partial retraining~\cite{FedEraser} —
but lack rigorous guarantees and often leave residual data traces in the model.
Current exact FU methods~\cite{tao2024communication, xiong2023exact, qiu2023fedcio}, such as \textit{FedCIO}~\cite{qiu2023fedcio}, extend the centralized SISA (Sharded, Isolated, Sliced, and Aggregated)~\cite{bourtoule2021machine} training framework to the distributed case.
They partition clients into disjoint clusters and train cluster-specific models independently.
Upon an unlearning request, only the cluster containing the deleted data must be retrained.
While this approach reduces retraining scope, its reliance on cluster retraining still incurs high communication costs and frequent service downtime, particularly when unlearning requests arrive sequentially.

To overcome the high retraining cost and service failure limitations, we draw inspiration from the recently proposed centralized \textit{Sequence-aware Sharded Sliced Training (S3T)} framework~\cite{chowdhury2024towards}, which integrates Parameter Efficient Finetuning (PEFT) to achieve parameter isolation.
S3T sequentially trains lightweight PEFT modules with distinct data partitions, enabling exact unlearning by simply deactivating the modules associated with deleted data.
Furthermore, S3T trains on multiple permutations (sequences) of the data slices to enhance robustness against multiple deletion requests.
Retraining is deferred until all sequence variants for a shard are exhausted, at which point S3T initiates retraining to recover model accuracy.

However, extending S3T to FL introduces challenges due to its distributed nature.
First, it is non-trivial to generalize the notion of a ``slice'' to FL.
A natural attempt is to treat a cluster in FedCIO as a slice. 
However, under Non-IID and uneven client data, it becomes difficult to form balanced data groups (i.e., clusters) for sequence-aware training across clients, which cannot only degrade accuracy, but also induce larger fluctuations in both the learning~\cite{li2022federated} and unlearning processes.
Second, developing theoretical models to compute the deletion rate (i.e., the expected number of deletions before retraining) and model performance are often required for exact model unlearning, because system designers must predict service availability under long-running deletion workloads and choose partition/sequence budgets without exhaustively tuning over many possible deletion patterns. However, it is particularly challenging in FL environments due to the interaction between grouping, sequencing, and stochastic unlearning.


\mypara{Contributions} 
In this paper, we propose \method, an exact federated unlearning framework. 
We propose to allow clients to partition their data and join multiple groups and dispatch local slices of data to these groups. 
Since the goal is to partition all clients' data into uniform groups, a client may participate in multiple groups by contributing local slices of data to each group to balance the data distribution, while its group membership (i.e., local slices) is constrained to keep communication cost under control. We formalize this problem and propose a randomized balanced grouping strategy and a Barycenter-aware centering grouping strategy to resolve the problem.
\method then trains multiple sequences of PEFT modules, each corresponding to a different group permutation. 
This design isolates the influence of different data groups into lightweight, independent modules, 
enabling instant exact unlearning - the server can 
simply deactivate modules corresponding to the unlearning data without retraining. 
Furthermore, using multiple training sequences with our designed sequence selection method, \method maintains high model utility even as deletion requests accumulate.
To complement the system design, we also develop a theoretical analysis leveraging probabilistic models of sequential deletions and dependency-aware training structures. It provides principled guidance for parameter selection and demonstrates \method's service-maintenance advantage and low communication overhead over alternative exact unlearning baselines.

Our primary contributions are summarized as follows:

\begin{itemize}
    \item We propose \method, a federated exact unlearning framework that efficiently partitions client data into groups mapped to lightweight PEFT modules. Each client contributes to a limited number of groups, achieving exact unlearning through instant module deactivation while keeping the training-phase communication cost bounded.

    \item We provide a rigorous theoretical analysis of the performance and efficiency for both \method and existing baselines. This includes analyzing their deletion rate, the expected model performance, and the training efficiency (i.e., communication overheads). 
    
    \item We conduct extensive experiments demonstrating that \method achieves comparable learning performance and training efficiency to baselines, while ensuring significantly longer service maintenance cycles under continuous unlearning requests, where baselines suffer from early service failure.
    We also conduct extensive ablation studies to validate the robustness of our method.
\end{itemize}

\section{Preliminaries}

\mypara{Machine Unlearning}
Machine unlearning is broadly classified into two categories: approximate and exact unlearning. 
Approximate unlearning~\cite{gupta2021adaptive,lin2023erm, zhang2024contrastive} aims to efficiently modify a trained model's parameters to approximate a state as if the target data had never been included. This is often achieved through methods like gradient ascent or by using designated retention and forget sets. 
However, while efficient, the stochastic nature of this process makes it difficult to audit and may provide weak privacy guarantees.

Exact unlearning~\cite{bourtoule2021machine, chen2022recommendation, yan2022arcane}, in contrast, provides a formal guarantee that the data's influence is completely removed by modifying the original training process. 
The seminal work in this area is SISA~\cite{bourtoule2021machine} (Sharded, Isolated, Sliced, and Aggregated training). 
SISA trains an ensemble of models, each on a disjoint data shard. 
To reduce costs, each shard is further divided into slices, and the model is trained incrementally, saving checkpoints after each slice. Upon a deletion request, SISA achieves exact unlearning by reverting to the checkpoint before the affected slice and retraining only on the remaining data within that shard.
However, SISA's single sequential model is fragile; deleting an early slice can invalidate all subsequent checkpoints, causing a catastrophic ``service failure" that requires a costly full retrain. 
S3T~\cite{chowdhury2024towards} addresses this issue by training multiple models on different permutations (sequences) of the data slices within a given budget. When a deletion request invalidates one sequence, S3T ensures longer service maintenance by instantly switching to the best-performing model from the remaining, still-valid sequences.

\mypara{Federated Unlearning}
Prior works on federated unlearning (FU) can be broadly categorized as approximate or exact. Approximate FU methods~\cite{liu2021federaser, wu2022federated, cao2023fedrecover, FUPGA, wu2022federated, pan2025federated} attempt to efficiently estimate the unlearned model. 
Some approaches, like FedEraser~\cite{liu2021federaser}, are history-dependent, utilizing stored historical updates to approximate the model state post-deletion. Other methods, such as FUPGA~\cite{FUPGA,pan2025federated}, are gradient-based, attempting to reverse a client's contribution by applying gradient ascent. 
However, these approximate methods have notable limitations. 
History-dependent approaches incur impractical storage and communication overhead, as they must save numerous client updates or model checkpoints. Gradient-ascent techniques risk gradient conflicts with remaining data, which can destabilize the global model and degrade its utility, especially in Non-IID settings. 
Most importantly, neither approach provides a formal guarantee of complete data removal, which is often a strict requirement.

Exact FU methods~\cite{tao2024communication, xiong2023exact, qiu2023fedcio}, in contrast, provide a verifiable guarantee that the data's influence is completely removed in FL. 
A state-of-the-art example is FedCIO~\cite{qiu2023fedcio}, which adapts the SISA framework to the federated setting. 
FedCIO's strategy involves partitioning clients into disjoint clusters, training an independent model for each cluster in isolation, and performing a one-shot aggregation to form the global model. 
While FedCIO achieves exactness, its limitation is its unlearning mechanism: upon a deletion request, it must retrain the entire affected cluster from scratch. 
This ``retrain-from-scratch" paradigm is computationally prohibitive, incurs high latency, and leads to the catastrophic service downtime that our work seeks to resolve. In addition, the accuracy and learning stability of FedCIO will degrade with Non-IID data since its grouping of clients by gradient similarity cannot ensure balanced groups.

\mypara{Federated Parameter-Efficient Fine-Tuning}
The ever-increasing size of large-scale models makes direct fine-tuning prohibitively expensive. To mitigate this, Parameter-Efficient Fine-Tuning (PEFT) methods have been proposed. These approaches introduce a small number of additional trainable parameters, while keeping the vast majority of the pre-trained parameters frozen. Among these methods, LoRA \cite{hu2022lora} is arguably the most popular, achieving performance comparable to full fine-tuning while training less than 1\% of the total parameters.

While PEFT addresses the computational cost, the data required for fine-tuning is often domain-specific, siloed across multiple parties, and cannot be shared directly due to privacy concerns. 
Recent works propose Federated PEFT
\cite{zhang2022federated, kuang2024federatedscope} for reducing communication and computation cost in FL by only training, communicating and aggregating the small set of PEFT parameters (e.g., LoRA matrices) instead of the entire model.
We include a broader discussion of related work in~\autoref{appendix:related_work}.

\section{Problem Statement and Analysis}
\begin{table}[!tbp]
\centering
\caption{Summary of notations.}
\label{tab:notation}
\begin{tabularx}{\columnwidth}{@{} l >{\raggedright\arraybackslash}X @{}} 
\toprule
\textbf{Notation} & \textbf{Description} \\
\midrule
$L$            & Total number of groups in \method \\
$c$            & Total number of clusters in FedCIO \\
$B$            & Training budget (number of permutations) \\
$B'$           & Effective budget, $\min(L, B)$ \\
$G_i$        & The $i$-th data group in \method \\
$\mathbb{P}$ & Probability distribution of models\\
$N$            & Clients number \\
$\mathcal{D},|\mathcal{D}|$  & Total training data and datasize \\
$\delta(\cdot)$ & Deletion Rate (expected requests until failure) \\
$H_n$          & $n$-th Harmonic Number $\sum_i^n\frac{1}{i}$ \\
$r$            & Number of unlearning requests (deletions) \\
$R$            & Remaining samples in the best sequence post-unlearning \\
$U$  & Cyclic Span (min. contiguous groups covering) \\
$T$   & Communication rounds in FL system \\
$S(r,m)$       & Stirling number of the second kind \\
\bottomrule
\end{tabularx}
\end{table}

In this section, we formalize the federated learning setup and the exact unlearning problem. 
We begin by introducing the problem formulation, including both the learning stage and the unlearning stage within a federated system.
We then outline the key design goals for an exact federated unlearning framework that preserves model utility, supports long-term service maintenance, and remains training-efficient.
Finally, we review existing federated unlearning solutions and highlight their limitations, particularly their inability to support multiple unlearning requests efficiently due to high communication and retraining costs.
These limitations motivate the need for a scalable and communication-efficient exact unlearning approach.
The main notations used in this paper are summarized in Table~\ref{tab:notation}.

\subsection{Problem Statement}

\mypara{Learning Stage}
We consider a federated fine-tuning setting where a central server coordinates a set of $N$ participating clients 
$\mathcal{C} = \{c_1, c_2, \dots, c_N\}$.  
Each client $c_i$ holds a local dataset $\mathcal{D}^i$, which is potentially Non-IID across clients.  
We focus on the fine-tuning setting where the global model is initialized with a pre-trained backbone
(e.g., a large vision or language model) 
and is collaboratively fine-tuned across clients to adapt to downstream tasks.  

To reduce communication and computation costs, only lightweight PEFT modules are trained and exchanged between clients and the server,  while the backbone parameters remain frozen and stored locally throughout the process.

\mypara{Unlearning Stage} 
Our goal is to achieve \textbf{exact unlearning} on the global model, ensuring that the unlearned data is not included in the training data of the updated model.
Let $\mathbf{M}$ denote the global model obtained after the learning stage, $\widetilde{\mathbf{M}}$ denote the global model after unlearning. 
Formally, for an unlearning request targeting a forget set $\mathcal{D}_f$, the unlearning operator $\mathcal{U}$ produces $\widetilde{\mathbf{M}}$ such that:
\[
    \mathbb{P}\bigl(\mathcal{U}(\mathcal{D}, \mathcal{D}_f, \mathbf{M})\bigr)
    = 
    \mathbb{P}\bigl(\mathcal{A}(\mathcal{D} \setminus \mathcal{D}_f)\bigr),
\]
where $\mathcal{A}$ is the original learning algorithm and $\mathcal{D} \setminus \mathcal{D}_f$ is the remaining dataset.

In an FL system, unlearning requests are triggered by clients—e.g., a client $c_i$ requests that their data $\mathcal{D}^i$ be removed in whole or in part. 
As such requests may arrive \emph{sequentially}, the server must incrementally update the global model in response to each request, ensuring that the current model does not retain any residual influence from the removed data. 
Moreover, the system must preserve service availability throughout the unlearning process: the time to service recovery after each deletion must be minimized to avoid extended downtime or performance degradation.

\mypara{Design Goals}
Given the above setting, we aim to develop an exact federated unlearning protocol with the following design goals:

\begin{itemize}
    \item \textbf{G1: Model Utility.}
    The framework aims for high model utility across the \textit{learning} and the \textit{unlearning} stages. 
    During learning, the global model $\mathbf{M}$ should minimize the expected loss over all client data. 
    During unlearning, the unlearned model $\widetilde{\mathbf{M}}$ must maintain comparable utility on the remaining dataset $\mathcal{D} \setminus \mathcal{D}_f$. 

    \item \textbf{G2: Service Maintenance.}
    The framework aims to reduce the service downtime.
    We should maximize the number of supported unlearning requests before a full retraining becomes necessary.
    Additionally, the communication or computation cost for each necessary retraining should be small.

    \item \textbf{G3: Training Efficiency.}
    The training process $\mathcal{A}$
    must introduce only modest computational and communication overhead, compared to standard federated learning with PEFT.
\end{itemize}

\subsection{Analysis of Existing Solutions}

\mypara{FedRetrain}
The most straightforward baseline is federated retraining (FedRetrain), where for each unlearning request, the server and all clients perform a complete retraining process on the remaining data. 
While this approach naturally satisfies \textbf{G1} (Model Utility) and avoids additional communication overhead, its major drawback lies in extremely high retraining cost and prolonged service downtime.
Thus, FedRetrain fails to meet and \textbf{G2} (Service Maintenance) and \textbf{G3} (Training Efficiency), as it cannot efficiently handle sequential unlearning requests or maintain continuous service availability.

\mypara{FedCIO}
For the FedCIO approach~\cite{qiu2023fedcio},  
during its unlearning stage, when a deletion request occurs, only the affected cluster needs to be retrained, while other cluster models remain available for inference.
Although this design improves \textbf{G3} (Training Efficiency), its gradient-based clustering of clients may degrade both training accuracy and stability and compromise \textbf{G1} (Model Utility). Moreover, it still suffers from substantial retraining time and fails to satisfy \textbf{G2} (Service Maintenance) once multiple clusters are affected.
Consequently, FedCIO experiences early service failure and cannot sustain long-term service maintenance without retraining.

\begin{figure} [!tbp]
\centering  
\includegraphics[width=0.35\textwidth]{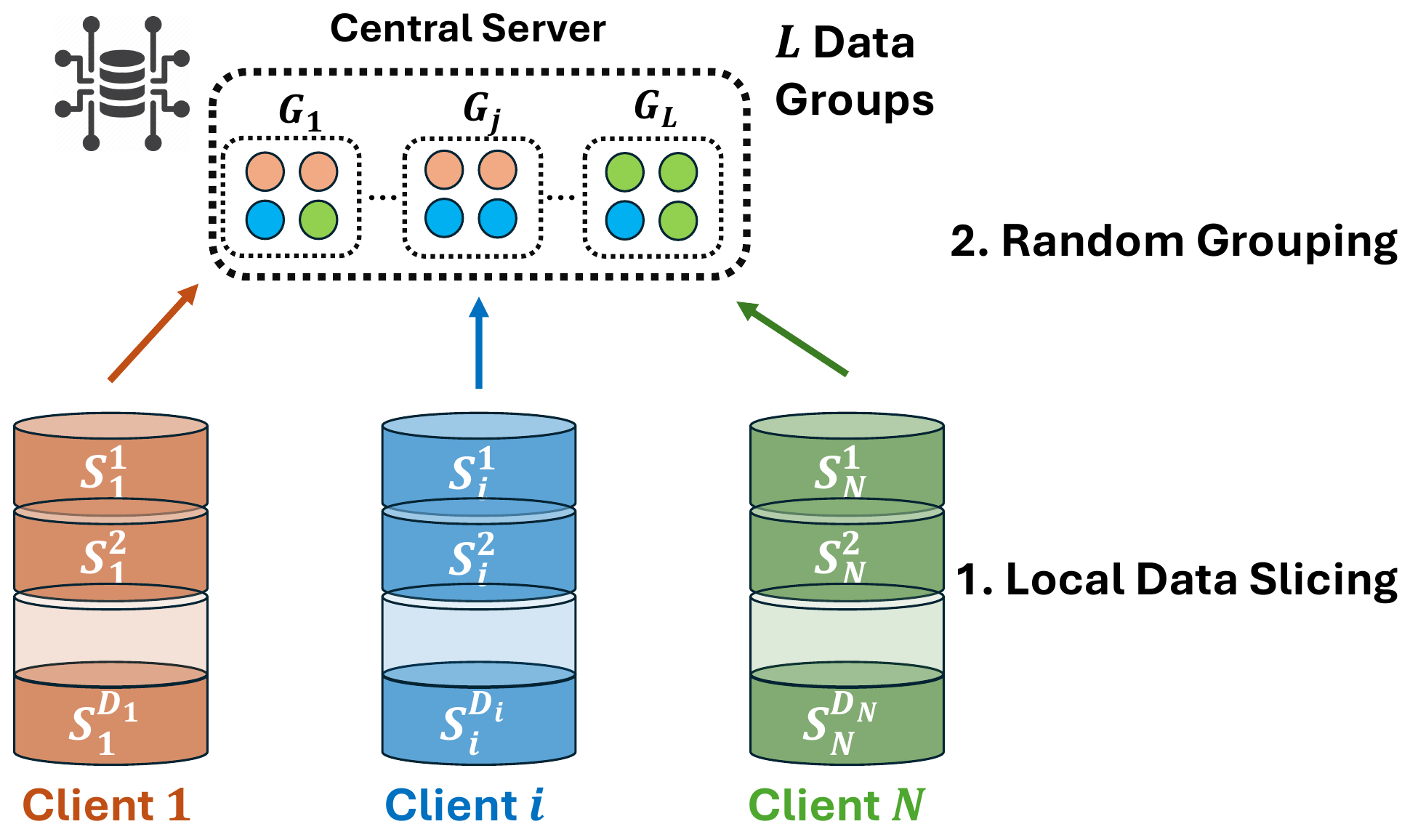}
\caption{Illustration of the federated data grouping method.
Each client locally slices its dataset into multiple pieces, and the central server randomly shuffles and assigns these slices into balanced groups for sequential group-based training.}
\label{fig:data_group}
\end{figure}

\section{\method}

We present \method (\underline{Fed}erated \underline{S}equential \underline{G}roup-based \underline{T}raining), an exact federated unlearning framework designed to meet all three design goals: \textbf{G1} (Model Utility), \textbf{G2} (Service Maintenance), and \textbf{G3} (Training Efficiency). 
The key insight is to achieve \textit{parameter isolation} through multiple lightweight model sequences, where each sequence corresponds to a distinct permutation of $L$ grouped client datasets. 
To efficiently realize this idea in a federated setting, \method leverages PEFT modules that are trained sequentially on these data groups. 

By maintaining these PEFT modules server-side and training them in a cumulative yet separable manner, \method isolates the contribution of each group into independent parameter subsets. 
This design enables exact unlearning by simply deactivating the PEFT modules associated with the deleted group—without immediate retraining. 
Consequently, the framework preserves model utility (\textbf{G1}), maintains long-term online service (\textbf{G2}), and incurs minimal communication and computation overhead(\textbf{G3}). 

In the following subsections, we detail the core components of \method: 
\autoref{subsec:grouping} introduces the data grouping strategy that partitions client data into balanced groups; 
\autoref{subsec:SGT} describes the sequential group-based training process, which trains PEFT modules along multiple ordered sequences; 
and \autoref{subsec:unlearning} presents the unlearning stage, which performs efficient and exact data removal through module deactivation.

\begin{figure*} [!tbp]
\centering  
\includegraphics[width=0.9\textwidth]{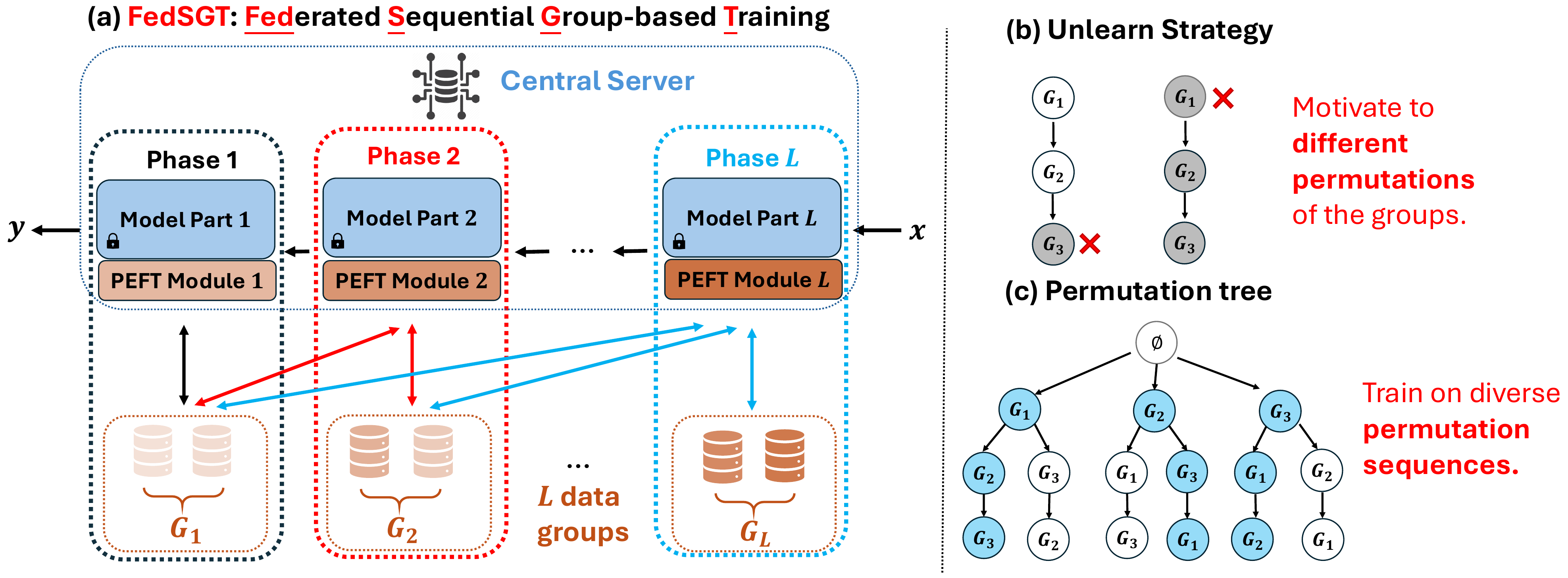}
\caption{(a) \textbf{\method overview}. 
Each phase incrementally trains a new PEFT module on the cumulative data groups 
$\{G_1,\dots,G_i\}$, while freezing previous modules. 
This design enables exact unlearning by isolating the contribution of each group.  
(b) \textbf{Unlearn Strategy:} 
If the earliest trained group (e.g., $G_1$) is removed, all subsequent modules that depend on it become invalid, motivating the need to diversify training orders.  
(c) \textbf{Permutation Tree:} 
To improve service maintenance, multiple group permutations are trained as independent sequences, 
distributing dependencies across different paths so that unlearning a single group only disables a subset of sequences while keeping others functional.
}
\label{fig:fedSGT}
\end{figure*}

\subsection{Data Grouping}
\label{subsec:grouping}

We consider the dataset partitioned among $N$ clients, each of which is split into $D_i$ local \emph{slices}. 
These $M = \sum_{i=1}^{N} D_i$ slices serve as the atomic training units in \method. 
The goal of the grouping stage is to organize these heterogeneous client slices into multiple balanced groups that facilitate both efficient learning and exact unlearning.


\mypara{Data Grouping Objective Formulation}
Given all client slices $\{S_{i,j}\}$, our goal is to partition them into $L$ groups $\{G_1,\dots,G_L\}$ such that each group remains representative of the overall data distribution. This is crucial for sustaining utility when groups are progressively removed by unlearning requests, when data is Non-IID.
Meanwhile, we enforce balanced group sizes for communication fairness and allow each client to participate in at most $B_i$ groups to bound edge-device workload.

We minimize the deviation between each group's summary statistics and the global statistics. Concretely, letting $\mu(G_g)$ denote the mean slice embedding within group $G_g$ and $\mu(\mathcal{D})$ the global mean, we consider:
\begin{align}
\min_{\{G_g\}_{g=1}^{L}} \quad & \sum_{g=1}^{L}\left\|\mu(G_g)-\mu(\mathcal{D})\right\|_2^2 \label{eq:group_obj_mu}\\
\text{s.t.}\quad 
& |G_g| \in \{\lfloor M/L\rfloor,\lceil M/L\rceil\}, \ \forall g \label{eq:group_balance_mu}\\
& \sum_{g=1}^{L}\mathbf{1}\!\left(\exists j:(i,j)\in G_g\right)\le B_i,\ \forall i. \label{eq:group_budget_mu}
\end{align}
This formulation is combinatorial due to the discrete assignment of slices to groups.

\mypara{Computational Hardness}
Problem~\eqref{eq:group_obj_mu} is NP-hard in general due to its discrete slice-to-group assignment nature.
Intuitively, even in a highly restricted setting with two balanced groups ($L=2$ and equal group sizes), minimizing the deviation between each group mean and the global mean becomes equivalent to finding an equal-cardinality partition whose two subset sums are as close as possible.
This is exactly the classic balanced partition problem, which is NP-hard~\cite{karp2009reducibility}.
Adding the per-client participation budget~\eqref{eq:group_budget_mu} further makes exact optimization impractical at FL scale and motivates lightweight heuristics.

\mypara{Our Solutions}
We therefore adopt efficient heuristics: a randomized balanced partition as the default grouping (\autoref{alg:random-grouping}) and a Barycenter-Aware Centering (BAC) as an alternative grouping strategy (detailed in \autoref{appendix:group}).

\mypara{Randomized Balanced Grouping}
Given the hardness above, our default strategy is a purely index-driven randomized balanced partition (\autoref{alg:random-grouping}): we permute all slice indices using a random seed and evenly split the permuted list into $L$ groups.
This yields (i) nearly equal group sizes; (ii) strong mixing across clients in expectation; and (iii) negligible overhead and no dependence on data semantics or model embeddings, avoiding additional privacy risks.
This choice is also supported by classical results in randomized load balancing and scheduling: simple randomization often achieves robust expected balance and can match or outperform more structured deterministic strategies, while avoiding NP-hard optimization at runtime~\cite{sitaraman2001power}.

\mypara{Barycenter-Aware Centering Grouping and Limitations}
We also consider an embedding-aware variant, Barycenter-Aware Centering (BAC), which greedily assigns slices to reduce $\|\mu(G_g)-\mu(\mathcal{D})\|_2$ more directly.
At a high level, BAC repeatedly selects a group with the largest barycenter deviation and inserts a slice that best decreases this deviation, under balance and budget constraints.
While BAC is more optimization-driven, it has notable limitations in our setting: it introduces extra computation and coordination, and may depend on model states or learned representations, raising additional privacy and systems concerns. 
Empirically, we find that the simpler randomized strategy already provides sufficiently good representativeness and robustness, often outperforming BAC once we account for runtime overhead and the inherent stochasticity of FL training.
Detailed comparisons are deferred to \autoref{appendix:group}.

\begin{algorithm}[!tbp]
\caption{Data Grouping of Client Slices}
\label{alg:random-grouping}
\textbf{Input:} Client set $\mathcal{C} = \{c_1,\dots,c_N\}$; per-client slices $\{S_{i,1},\dots,S_{i,D_i}\}$; number of groups $L$; random seed $s$\\
\textbf{Output:} Data groups $ \{G_1,\dots,G_L\}$ and mapping $(i,j)\mapsto \mathrm{group\_id}(S_{i,j})$\\
\textbf{Procedures:}
\begin{algorithmic}[1]
    \STATE $M \leftarrow \sum_{i=1}^N D_i$   \cmt{total number of slices}
    \STATE $\mathcal{I} \leftarrow \{(i,j) \mid i\in[1..N],\, j\in[1..D_i]\}$ \cmt{index pairs for all slices}
    \STATE $\pi \leftarrow \mathrm{Permute}(\mathcal{I}; s)$  \cmt{shuffle indices using random seed $s$}
    \STATE $q \leftarrow M / L$  \cmt{number of slices per group}
    \FOR{$g = 1$ \textbf{to} $L$}
        \STATE $G_g \leftarrow \{\pi[(g-1)q+1],\, \pi[(g-1)q+2],\, \dots,\, \pi[gq]\}$ \\ \cmt{assign $q$ slices to group $g$}
    \ENDFOR
    \FOR{\textbf{each} $(i,j) \in \mathcal{I}$}
        \STATE set $\mathrm{group\_id}(S_{i,j}) \leftarrow g$ \textbf{iff} $(i,j) \in G_g$
    \ENDFOR
\end{algorithmic}
\textbf{Return:} Group set $G = \{G_1,\dots,G_L\}$ and mapping $(i,j)\mapsto \mathrm{group\_id}(S_{i,j})$
\end{algorithm}


{\remark
Different from typical machine unlearning frameworks such as SISA~\cite{bourtoule2021machine}, the definitions of \textit{group} and \textit{slice} in our method are distinct.
In SISA, the dataset is first divided into multiple \textit{shards}, each of which is further partitioned into \textit{slices} trained sequentially.  
In contrast, our framework considers only a single global shard, while each group corresponds to an independent subset of data—analogous to a slice in SISA. 
}

\subsection{\underline{S}equential \underline{G}roup-based \underline{T}raining (SGT)}
\label{subsec:SGT}

After grouping, the federated training proceeds in a cumulative and sequential manner, as illustrated in \autoref{fig:fedSGT}.  
Let $\{G_1, G_2, \dots, G_L\}$ denote the $L$ grouped datasets.  
The central server maintains a shared backbone model and a series of PEFT modules $\{\mathbf{P}_1, \dots, \mathbf{P}_L\}$, each corresponding to one training phase.  

At Phase 1, the model is trained with the data from $G_1$, producing the first PEFT module $\mathbf{P}_1$.  
At Phase 2, training continues on the cumulative dataset $\{G_1, G_2\}$, initializing a new PEFT module $\mathbf{P}_2$ while keeping $\mathbf{P}_1$ frozen.  
This process repeats sequentially—at Phase $i$, the model is fine-tuned on $\{G_1, \dots, G_i\}$, generating the PEFT module $\mathbf{P}_i$.  
Hence, each module $\mathbf{P}_i$ encodes the incremental contribution of the $i$-th group beyond what was already captured by the preceding ones.

It is worth noting that each group $G_i$ may involve data slices from multiple clients.  
During the training of group $G_i$, the corresponding clients participate in a standard federated fine-tuning process where each client locally updates the current PEFT module on its local portion of $G_i$,  
and the central server aggregates their updates to obtain the global module parameters.  
Therefore, within each phase, the procedure is fully compatible with conventional federated training, while across phases, it progresses cumulatively through the sequence of groups.  

This design achieves two key properties:  
(1) it enables exact unlearning—removing a group $\mathcal{C}_i$ can be done by deactivating all PEFT modules $\{\mathbf{P}_i, \dots, \mathbf{P}_L\}$ that depend on it;  
and (2) it ensures training efficiency, as each phase updates only a small number of PEFT parameters while preserving previously learned knowledge.
The complete procedure is summarized in \autoref{alg:seq-training}.

\begin{algorithm}[!tbp]
\caption{Sequential Group-based Training}
\label{alg:seq-training}
\textbf{Input:} Grouped datasets $\{G_1, \dots, G_L\}$; PEFT backbone model $\mathbf{M}$; training budget $B$\\
\textbf{Output:} Trained model set $\mathcal{F} = \{\mathbf{M}_\rho \mid \rho \in \Pi\}$ \\[1pt]
\textbf{Procedures:}
\begin{algorithmic}[1]
    \STATE $\mathcal{F} \leftarrow \emptyset$  \cmt{initialize trained model set}
    \STATE $\Pi \leftarrow \textsc{SelectTrainingSequences}(L, B)$ \\ 
    \cmt{select up to $B$ group permutations $\rho$}
    \FOR{$\rho \in \Pi$}
        \STATE Initialize backbone model $\mathbf{M}_\rho \leftarrow \mathbf{M}$
        \FOR{$i = 1$ \textbf{to} $L$}
            \STATE $\mathbf{M}_\rho \leftarrow \textsc{GroupWiseTraining}(\mathbf{M}_\rho, \{G_{\rho(1)}, \dots, G_{\rho(i)}\})$ 
            \STATE Freeze $\text{PEFT module } \mathbf{P}_{\rho(i)}$ \\
            \cmt{preserve previous knowledge}
        \ENDFOR
        \STATE $\mathcal{F} \leftarrow \mathcal{F} \cup \{\mathbf{M}_\rho\}$  \cmt{store trained sequence model}
    \ENDFOR
\end{algorithmic}
\textbf{Return:} Trained model set $\mathcal{F} = \{\mathbf{M}_\rho \mid \rho \in \Pi\}$
\end{algorithm}

\mypara{Training with Multiple Sequences}
While the sequential group-based training described above enables exact unlearning, it still suffers from a key limitation: 
if the earliest trained group (e.g., $G_1$) is removed, all subsequent PEFT modules depend on it, and thus the entire model becomes invalid. 
As illustrated in \autoref{fig:fedSGT}(b), this dependency chain causes early service failure when unlearning requests affect the prefix of the sequence.

To mitigate this issue, we extend the training process to multiple \textbf{permutation sequences} of group orders, as shown in \autoref{fig:fedSGT}(c). 
Instead of relying on a single order $\{G_1, G_2, \dots, G_L\}$, the server trains multiple independent sequential models, each following a distinct permutation of the groups. 
Each sequence defines an alternative dependency path, enabling the system to recover valid model variants even after certain groups are deleted.

This multi-sequence design substantially improves \textbf{service maintenance} by increasing the number of supported unlearning requests before retraining becomes necessary. 
Training on diverse group permutations distributes dependency structures across multiple paths, ensuring that the removal of any individual group disables only a subset of sequences while keeping others fully functional.

However, introducing more training sequences inevitably increases computational and communication overhead.  
To balance service robustness and efficiency, we define a training budget $B$ that limits the maximum number of sequences trained.  
Within this budget, the server selects a subset of representative group permutations for sequential training to achieve optimal coverage of unlearning scenarios.

\begin{table*}[!t]
\centering
\setlength{\tabcolsep}{4pt} 
\renewcommand{\arraystretch}{1.5} 
\caption{
Sequence and model selection results under different sequence selection strategies at three cumulative unlearning stages. 
The system is configured with $G=6$ groups and a training budget of $B=6$. 
Red highlights (\textcolor{red}{\texttt{...}}) indicate the remaining active PEFT modules after each unlearning.
}
\label{tab:seq-example}
\begin{tabular}{c|c|c|c}
\specialrule{.1em}{0pt}{0pt}
\textbf{Strategy} 
& \textbf{Time 1 (unlearning on $\{G_1\}$)} 
& \textbf{Time 2 (unlearning on $\{G_5\}$)} 
& \textbf{Time 3 (unlearning on $\{G_2\}$)} \\
\specialrule{.06em}{0pt}{0pt}
AllSeq  
& $[\textcolor{red}{0}12345], [\textcolor{red}{50}1234], [\textcolor{red}{450}123], [\textcolor{red}{3450}12], [\textcolor{red}{23450}1]$ 
& $[\textcolor{red}{0}12345], [\textcolor{red}{4}50123], [\textcolor{red}{34}5012], [\textcolor{red}{234}501]$
& $[\textcolor{red}{0}12345],[\textcolor{red}{4}50123],[\textcolor{red}{34}5012]$ \\
\hline
MinSeq  
& $[\textcolor{red}{23450}1]$  
& $[\textcolor{red}{0}12345], [\textcolor{red}{234}501]$  
& $[\textcolor{red}{0}12345],[\textcolor{red}{34}5012]$ \\
\hline
LongSeq  
& $[\textcolor{red}{23450}1]$  
& $[\textcolor{red}{234}501]$  
& $[\textcolor{red}{34}5012]$ \\
\specialrule{.1em}{0pt}{0pt}
\end{tabular}
\end{table*}

\subsection{Unlearning}
\label{subsec:unlearning}

After training, each sequence model $\mathbf{M}_\rho$ corresponds to a distinct dependency path among group datasets $\{G_1, \dots, G_L\}$. 
When an unlearning request arrives for a specific group $G_u$, the system performs exact unlearning by \textbf{deactivating} all PEFT modules that depend on this group.  
Concretely, for any sequence $\rho \in \Pi$, if $G_u$ appears at position $k$ in $\rho$, then all subsequent modules 
$\{\mathbf{P}_{\rho(k)}, \mathbf{P}_{\rho(k+1)}, \dots, \mathbf{P}_{\rho(L)}\}$ in that sequence are disabled, 
yielding a truncated yet valid model $\mathbf{M}^{(-u)}_\rho$.
We denote by $\mathbf{M}^{(-u)}_\rho$ the truncated model obtained from $\mathbf{M}_\rho$ 
after deactivating all PEFT modules that depend on group $G_u$.
This mechanism ensures that the remaining model contains no contribution from the deleted group, achieving exact and certifiable unlearning without retraining.

However, after certain groups are removed, different sequence models may remain partially functional.  
Since each sequence covers a distinct subset of the dependency paths, the system must determine which available model to use for inference.  
This introduces the challenge of sequence selection: given a deletion set $\mathcal{U}$, 
we must select from the remaining available sequence models $\{\mathbf{M}^{(-\mathcal{U})}_\rho\}$ to ensure the best utility and stability during inference.

\mypara{Sequence Selection Method}
After certain groups are unlearned, different sequence models retain different numbers of active PEFT modules.  
Hence, not all remaining models contribute equally to inference.  
To ensure stable prediction quality, we introduce a sequence selection mechanism that determines which available model(s) should be used for service and how to combine them.
We design three representative strategies:

\begin{itemize}
    \item \textbf{AllSeq:} Aggregate predictions from all available sequences through weighted averaging.  
    The weight of each sequence is proportional to the number of its currently active modules, 
    reflecting the extent of learned knowledge still preserved after unlearning.
    
    \item \textbf{MinSeq:} Select the sequence whose active modules share the least overlap with the deleted groups across all trained permutations.  
    When multiple sequences remain partially valid, this strategy favors the one most independent from the removed data, thereby minimizing potential interference.

    \item \textbf{LongSeq:} Select the single sequence with the longest active prefix (i.e., the largest number of remaining modules).  
    This strategy prioritizes the sequence with the deepest training depth, which typically preserves the best utility.
\end{itemize}

This weighted treatment naturally reflects the varying reliability among the remaining sequences: 
those retaining more activated modules contribute more to the final inference, 
while truncated sequences have a weaker influence.  
In \autoref{subsec:theory_performance}, we provide a detailed analysis of the sequence selection strategies and their impact on overall model utility.

{\example
We illustrate how different sequence selection strategies behave under consecutive unlearning requests.
Consider the \method system consists of $G=6$ data groups and $B=6$ trained sequences: 
$[012345], [501234], [450123], [345012], [234501], [123450]$.  
We simulate three unlearning stages, where the requests sequentially target group data 
$\{G_1\}$, $\{G_5\}$, and $\{G_2\}$.  
The resulting sequence and model selections are summarized in \autoref{tab:seq-example}.  
In general, the available sequences and models under MinSeq form a subset of those under AllSeq, while LongSeq further selects the longest surviving one from the MinSeq candidates.}



\section{Theoretical Analysis}
In this section, we present a theoretical analysis of \method and the baselines (FedCIO, FedAvg). 
We begin by formalizing the deletion rate for exact unlearning systems, then analyze utility during the unlearning stage and the associated overhead. 
Building on these theoretical results, we provide a direct comparison between \method and FedCIO to clarify their relative efficiency–utility trade-offs.

\subsection{Deletion Rate}
\label{subsec:theo_deletion_rate}

\begin{definition}[Deletion Rate~\cite{chowdhury2024towards}]
The deletion rate, $\delta(S)$, of an exact unlearning system $S$,
is the expected number of deletion requests the system can handle before a full retraining becomes necessary,
\textit{i.e.}, when no unaffected model remains available for service.
\end{definition}

The deletion rate quantifies a system’s tolerance to unlearning requests. The theoretical results are presented in the main text, with detailed proofs provided in~\autoref{sec:appendix_deletionrate}.

\mypara{\method}
In \method, when the budget $B>1$ , the system exhibits greater resilience to unlearning requests. 
Given $L$ total number of groups and a training budget $B$, we have $B'=\min(L, B)$ different groups at the topmost position in a sequence. 
The deletion rate quantifies the number of unlearning requests that can be processed before all $B'$  groups are affected, and is given by
\begin{equation*}
  \delta(\method) \approx L \cdot H_{B'}.  
\end{equation*}
where $H_{B'}$ is the $B'$-th harmonic number.

\mypara{FedCIO}
In FedCIO, complete failure occurs when all clusters are affected. 
Assuming all clusters are of equal size and each unlearning request randomly targets one cluster, the process reduces to a coupon collection problem \cite{blom20047}.

Let $c$ denote the number of clusters; the expected deletion rate for FedCIO is
\[
\delta(\text{FedCIO}) \approx c \cdot H_c,
\]
where $H_c$ is the $c$-th harmonic number.

Deletion rate grows with system degrees: \method\ increases with the budget \(B\) (via \(B'=\min(L,B)\)) and group count \(L\), while FedCIO increases with the number of clusters \(c\); both follow harmonic growth.

\subsection{Performance Guarantee}
\label{subsec:theory_performance}

\begin{figure} [!tbp]
\centering  
\includegraphics[width=0.7\columnwidth]{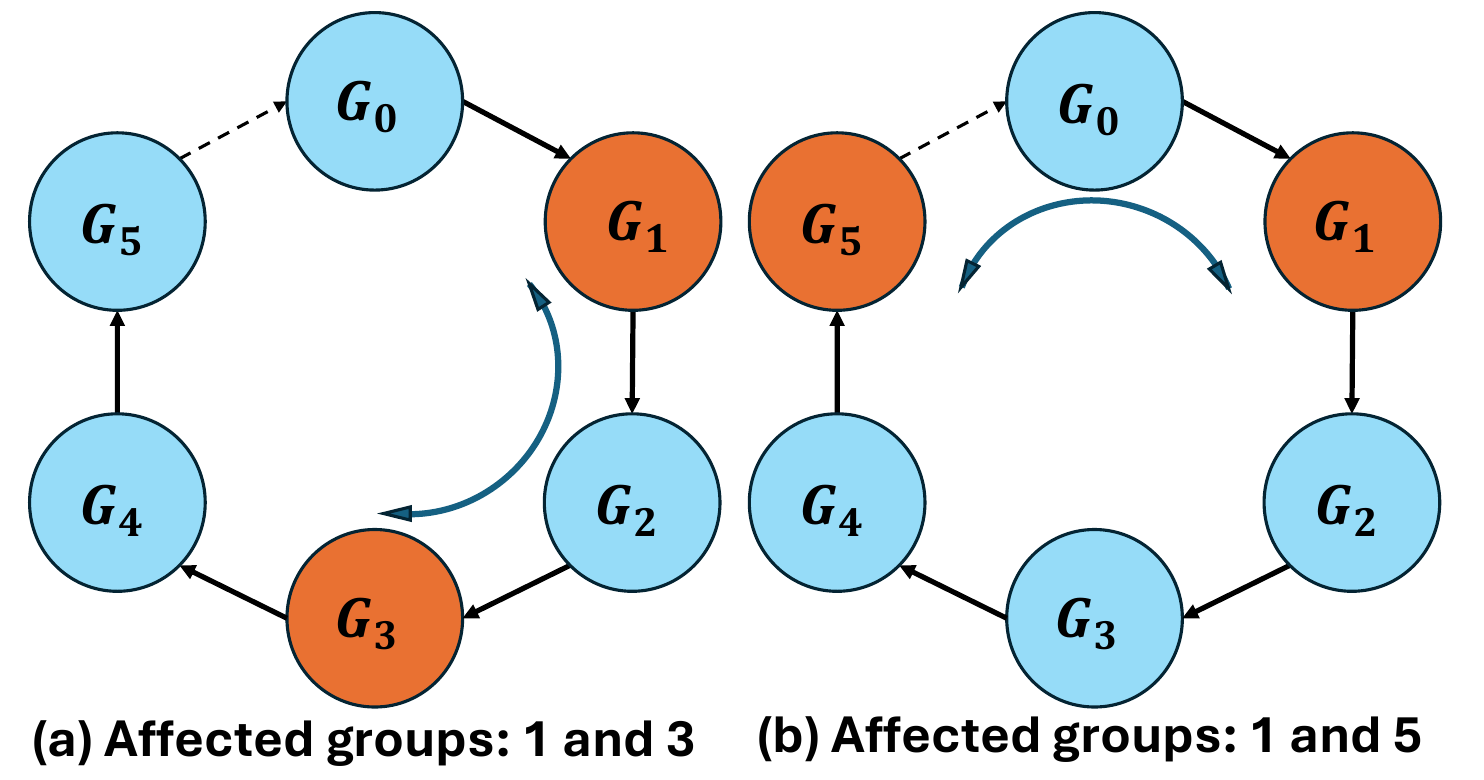}
\caption{Counting cyclic span. (a) If the affected groups are 1 and 3, the cyclic span is 3 because the minimum clusters to be removed are group 1, 2, and 3; (b) If the affected groups are 1 and 5, the cyclic span is also 3 because the minimum groups to be removed are 5, 0, and 1.
}
\label{fig:cyclic_span}
\end{figure}

While direct quantification of model performance without evaluation is challenging, we employ the expected number of remaining training samples 
$R$ associated with remaining PEFT modules after $r$ unlearning requests as a proxy for performance, given their positive correlation.

\mypara{\method}
Due to space constraints, we present only the analytical results in the main text. Detailed analyses and derivations are provided in~\autoref{sec:appendix_performance}.

\method employs a cyclic rotation algorithm to create $B>1$ sequences for each shard. It is difficult to get $\mathbb{E}_{B<L}[R]$ because one needs to analyze many different situations separately. Here we show how to get $\mathbb{E}_{B=L}[R]$, which is also our default setting in our experiments. Since the sequences when $B>L$ include every sequence when $B=L$ when using the cyclic rotation agorithm, so $\mathbb{E}_{B>L}[R]>\mathbb{E}_{B=L}[R]$. 

Since \method involves $B$ sequences coupled with different serving strategies (i.e., sequence selection strategies), we cannot simply use the total remaining training samples of all sequences to estimate the model performance. Thereby, we analyze the case for the LongSeq strategy, computing the expectation of the greatest number of remaining training samples in one sequence. It is given by
$\mathbb{E}[R]=\frac{|D|}{L}(L-\mathbb{E}[U])$, where $U$ is the \textit{cyclic span}, defined as the minimum number of slices to be removed in a sequence using the cyclic rotation algorithm with $B\ge L$. 
Figure~\ref{fig:cyclic_span} illustrates two cases of counting cyclic span. The expression for $\mathbb{E}[U]$ is
\begin{equation*}
\mathbb{E}[U]=\sum_{m=1}^{\min(r,L)} \Pr(M=m) \cdot \mathbb{E}[U\mid M=m].
\end{equation*}
The first term in the summation is given by
\begin{equation*}
    \Pr(M=m)=\binom{L}{m} \cdot m! \cdot S(r,m)/L^r
\label{eq:Pr(M=m)}
\end{equation*}
where $S(r,m)$ is the Stirling number of the second kind.
The second term in the summation is given by
\begin{align*}
    \mathbb{E}[U\mid M=m]  &=1+\sum_{s=1}^{L-1} \frac{1}{\binom{L-1}{m-1}} \times\\
&\sum_{j=0}^{\left\lfloor\frac{L-m}{s}\right\rfloor}(-1)^j\binom{m}{j}\binom{L-1-j s}{m-1}
\end{align*}

\mypara{FedCIO}
Let $X_i$ be the indicator that group $i$ is not affected after $r$ independent uniformly distributed unlearning requests. $P(X_i=1)=(1-\frac{1}{c})^r$, so by linearity of expectation, the expectation of number of groups that are unaffected after $r$ unlearning requests is $\mathbb{E}[\sum_{i=1}^c X_i]=\sum_{i=1}^c\mathbb{E}[X_i]=c(1-\frac{1}{c})^r$. The cluster size is $|D|/c$, therefore, the size of the training samples that remain is $\mathbb{E}[R]=\frac{|D|}{c} \cdot c(1-\frac{1}{c})^r= |D|(1-\frac{1}{c})^r$.

\subsection{Overhead Analysis}
\label{subsec:overhead}

\mypara{Assumptions}
Clients have equal-sized datasets; we set the number of data groups equal to the number of PEFT modules $L$.
We compare \method\ with FedAvg and FedCIO in training and communication.
With the backbone frozen, cost comes from (i) training PEFT modules and (ii) transmitting their updates.

\subsubsection{Training Complexity}
\label{subsec:train_complexity}
Let \(N\) be the number of clients, \(E\) local epochs, \(T\) rounds, \(|\mathcal D|=\sum_{i=1}^N|\mathcal D_i|\) the total data size, and \(P\) the trainable parameters per PEFT adapter.
Updating one module on \(|\mathcal D_i|\) costs \(O(|\mathcal D_i|P)\).

\mypara{FedAvg}
All clients fine-tunes a single PEFT module for \(E\) epochs over \(T\) rounds; across \(L\) modules the total training cost is
\[
O(T \cdot E \cdot |\mathcal{D}| \cdot P \cdot L).
\]

\mypara{FedCIO}
FedCIO partitions the \(N\) clients into \(C\) disjoint clusters and runs federated fine-tuning within each cluster for \(T\) rounds and \(E\) local epochs. 
Let \(|\mathcal D_k|\) be the data size of cluster \(k\) with \(\sum_{k=1}^C|\mathcal D_k|=|\mathcal D|\).
The per-cluster cost is \(O(TE|\mathcal D_k|P)\); summing over clusters (and \(L\) modules) yields
\[
O\!\left(\sum_{k=1}^C TE|\mathcal D_k|P \cdot L\right)=O\!\left(TE|\mathcal D|PL\right),
\]
matching FedAvg in overall compute.

\mypara{\method}
Unlike the baselines, \method\ trains sequential group-based modules: at phase \(i\), a new PEFT module \(\mathbf{P}_i\) is trained on \(\{G_1,\ldots, G_i\}\) while previous modules are frozen. Each phase uses \(E\) local epochs and one communication round; only one module is updated at a time, so the total number of rounds does not grow with \(L\).

Let \(|G_i|\) be the size of group \(G_i\) with \(\sum_{i=1}^{L}|G_i|=|\mathcal D|\).
The per-sequence cost sums \(L\) incremental stages:
\[
C_{\text{seq}}
= O\!\left(\sum_{i=1}^{L} E \big(\sum_{j=1}^{i} |G_j|\big) P\right)
\approx O\!\left(E\,\tfrac{L+1}{2}\,|\mathcal D|\,P\right),
\]
assuming balanced groups \(|G_i|\approx |\mathcal D|/L\).
With a budget of \(B\) sequences, the total training cost scales linearly:
\[
C_{\text{train}} = B\,C_{\text{seq}}
= O\!\left(BE\,\tfrac{L+1}{2}\,|\mathcal D|\,P\right).
\]
Thus, \method\ adds linear computation in \(B\) without extra communication rounds, trading modest overhead for stronger unlearning robustness and service continuity.

\mypara{Discussion} 
In \method, each phase uses a single communication round while keeping the same local epochs as the baselines ($E'=E$).
To match the total training cost of FedAvg/FedCIO, equate $C_{\text{train}}$:
\[
B \cdot E \cdot \frac{L+1}{2}
\;\approx\;
T \cdot E \cdot L
\Longrightarrow
\boxed{~ B \;\approx\; \frac{2\,T\,L}{L+1} ~}
\]

\subsubsection{Communication Overhead}

We evaluate communication overhead from the perspective of how many communication rounds each client participates in on average.  
This metric better reflects the true client-side cost, since different methods involve clients in different subsets of rounds.

\mypara{FedAvg}
In FedAvg, every client participates in every global communication round.  
Thus, each client takes part in exactly
\[
C^{\text{client}}_{\text{FedAvg}} = T
\]
rounds, where $T$ is the total number of FL rounds.

\mypara{FedCIO}
FedCIO adds an initial clustering stage requiring $T_{\text{cluster}}$ rounds, where all clients must participate to compute similarity statistics.  
After clustering, each cluster trains independently for $T$ rounds, but clients still participate in all communication rounds within their own cluster.

Since every client belongs to exactly one cluster, the per-client communication cost becomes
\[
C^{\text{client}}_{\text{FedCIO}}
= T_{\text{cluster}} + T.
\]

\begin{figure} [!tbp]
\centering  
\includegraphics[width=0.38\textwidth]{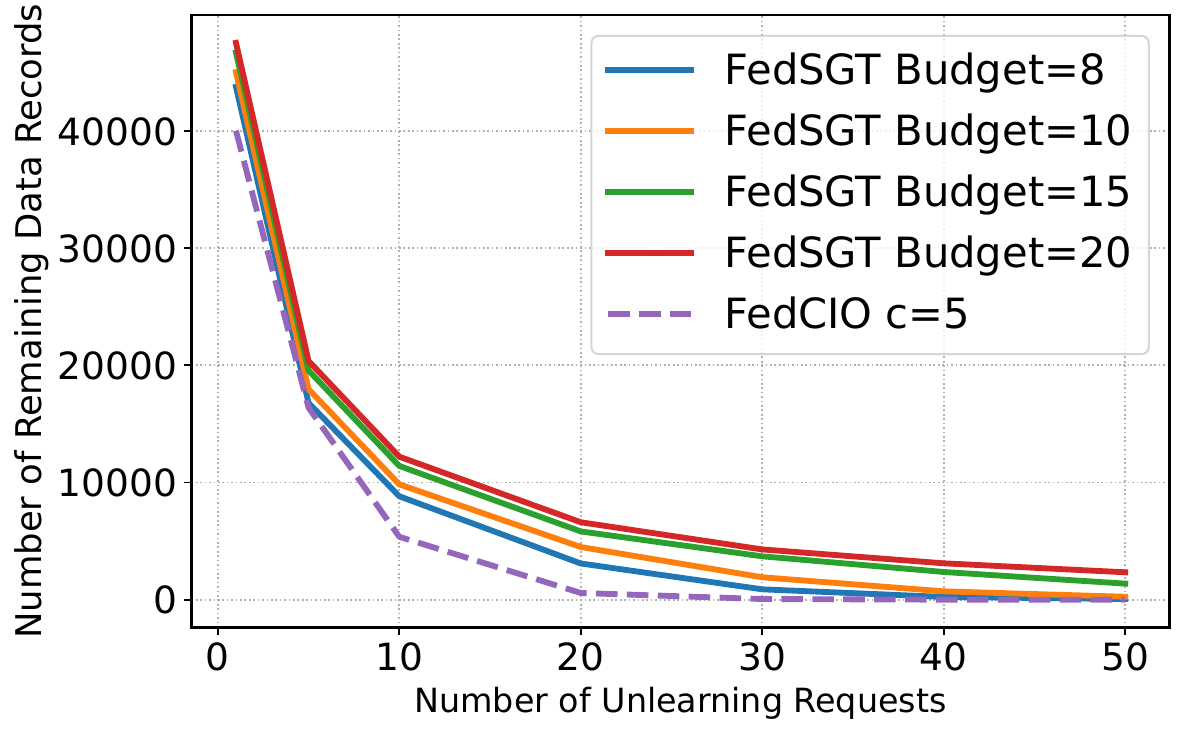}
\caption{Simulation results of performance analysis}
\label{fig:theoritical_analysis}
\end{figure}

\mypara{\method}
In \method, a client's communication cost depends on how many groups it participates in during the $L$ cyclic training sequences.
Let $K$ denote the number of groups that contain at least one of its $S$ slices after applying the uniform random grouping in~\autoref{alg:random-grouping}.  
Conditioned on $K=k$, we show in~\autoref{app:comm-derivation} that the client's total communication cost is
\[
\mathbb{E}\!\left[C_{\text{total}}(L)\mid K=k\right]
= L(L+1)\frac{k}{k+1}.
\]
Under random grouping, $K$ is itself a random variable whose distribution depends on $S$.
Combining these two components yields the exact expected communication cost:
\begin{equation}
\mathbb{E}[C_{\text{total}}(L,S)]
= L(L+1)
\sum_{k=1}^{\min\{S,L\}}
\Pr(K=k)\,\frac{k}{k+1},
\label{eq:main-comm}
\end{equation}
where $\Pr(K=k)$ is characterized in~\autoref{app:comm-derivation}.
Equation~\eqref{eq:main-comm} provides a closed-form expression for communication overhead as a function of the slice granularity $S$.

Considering the heterogeneity of edge devices in federated learning~\cite{gao2023fs}, we further analyze how the slicing granularity affects the performance--overhead trade-off, with detailed results provided in~\autoref{appendix:analysis_edge}.

\subsection{Comparison with Baselines}
We compare \method with FedCIO under matched training overhead.  
Consider a 10-client FL system with total data size 50{,}000, where \method uses $L=10$ groups and 2 slices per client, and FedCIO uses $c=5$ clusters.

\mypara{Matching training cost}
From Section~\ref{subsec:train_complexity}, if FedCIO uses $T=10$ training rounds (plus $T_{\text{cluster}}=2$ clustering rounds), the corresponding budget for \method is $B\!\approx\!36$.  
In our comparison, we evaluate smaller budgets $B\!\in\!\{8,10,15,20\}$ to study cost–performance tradeoffs.

\mypara{Unlearning utility and deletion rate}
Using the theoretical results (\autoref{subsec:theory_performance}), 
\autoref{fig:theoritical_analysis} illustrates the expected utility under sequential unlearning requests.  
As shown, \method consistently outperforms FedCIO across nearly all requests.  
This aligns with the theoretical deletion rates in~\autoref{subsec:theo_deletion_rate}:  
$\delta(\method)\!\approx\!29.29$ versus $\delta(\text{FedCIO})\!\approx\!11.42$,  
meaning \method is expected to maintain service for over $2.5\times$ more deletions.

\mypara{Training and communication overhead}
Under the setting, FedCIO performs $12$ communication rounds, whereas \method executes $71.5$ rounds (for $B=10$,$S=2$).  
Although this is more rounds ($\sim\!5.9\times$), each FedCIO round transmits all $L$ PEFT modules, while \method transmits only the active group’s modules (often just one or two).  
Consequently, the total communication bandwidth of the two methods remains comparable.

\mypara{FedCIO Cluster Numbers}
Although FedCIO’s deletion rate improves with more clusters, large $c$ values severely degrade model utility~\cite{qiu2023fedcio}, especially under Non-IID data, where training collapses toward per-client learning.  
We validate this effect experimentally in the next section.  
Thus, we adopt $c=5$ here as a practical balance between utility and cluster isolation.

\mypara{Summary}
Under approximately matched training and communication overhead,  
\method provides strictly higher unlearning utility and a substantially better deletion rate than FedCIO.  
The next section empirically validates these theoretical findings.




\section{Experiments}
In this section, we conduct extensive experiments to demonstrate and analyze \method's performance, compared to representative baselines. 


    
    
    

\subsection{Experiment Setup}

\mypara{Competitors}
We compare \method\ with two representative exact unlearning baselines:

\begin{itemize}
    \item \textbf{FedRetrain.}  
    The most straightforward exact unlearning baseline that fully retrains the federated model from scratch after each unlearning request.  
    While it guarantees correctness, it incurs the highest computation and communication cost.

    \item \textbf{FedCIO.}  
    A state-of-the-art exact FU that extends the centralized SISA framework by dividing clients into disjoint clusters and training each cluster independently.
\end{itemize}

\begin{figure*}[!tbp]
\centering
\subfigure[\normalsize Training performance under IID setting]{
\includegraphics[width=0.4\textwidth]{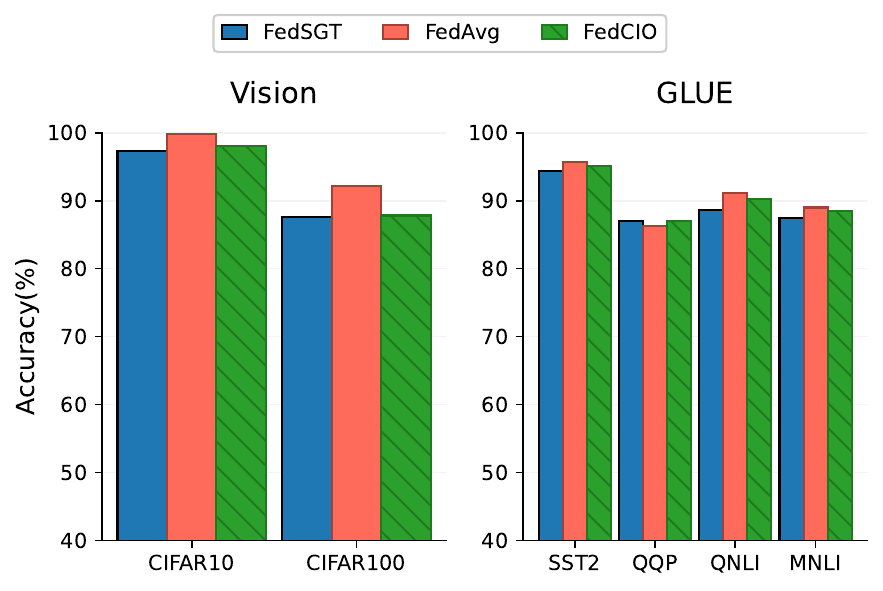}
\label{fig:train_results_iid}
}\hspace{0.5em}
\subfigure[\normalsize Training performance under Non-IID setting]{
\includegraphics[width=0.4\textwidth]{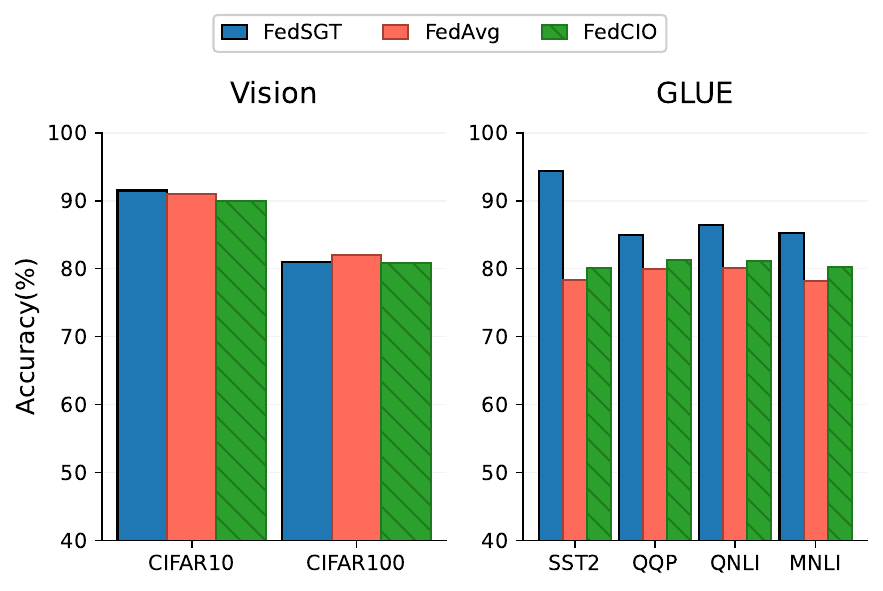}
\label{fig:train_results_non_iid}
}
\caption{Training performance comparison between \method, FedAvg, and FedCIO under IID and Non-IID settings.}
\label{fig:train_performance}
\end{figure*}

\mypara{Datasets and Models}
We evaluate \method\ on both vision and language model tasks to comprehensively assess its generality.  
For vision benchmarks, we use CIFAR-10 and CIFAR-100~\cite{krizhevsky2009learning}, adopting a $\text{ViT}_{\text{BASE}}$~\cite{dosovitskiy2020image} backbone for federated fine-tuning.  
For natural language understanding (NLU) tasks, we employ four datasets from the GLUE~\cite{wang2018glue} benchmark—SST-2, QQP, QNLI, and MNLI—using $\text{RoBERT}_{\text{LARGE}}$~\cite{liu2019roberta} as the base model.

\mypara{Implementation Details}
The experimental setup adopts the following hyperparameter configuration.  
We consider a federated system with 10 clients, 
for FedCIO, the number of clusters $c$ is set to 5.
The local training epochs are set to $3$ and $5$ for vision and language tasks, respectively.  
Both FedAvg and FedCIO are trained with $10$ communication rounds,  
while \method\ adopts a default training budget of $B=10$, and local slice data numbers $S$ = 5 and AllSeq unlearning sequence selection strategy.  
As analyzed in~\autoref{subsec:overhead}, this configuration corresponds to a lower overall training cost compared to the baselines.
Both \textit{IID} and \textit{Non-IID} settings are evaluated: the Non-IID distribution is simulated using a Dirichlet allocation with concentration parameter $\alpha = 0.3$. 
For the PEFT backbone, we use LoRA~\cite{hu2022lora} by default due to its broad applicability, efficiency, and ease of deployment across different model architectures. 
We also evaluate the impact of other PEFT methods on \method, and provide the detailed results in~\autoref{appendix:other_peft_ablation}.
We provide the details of the experimental setup and hyperparameters in~\autoref{appendix:exp_setup}.
The experiments were performed on a server equipped with NVIDIA H200 GPUs and 1 TB of system memory.

\begin{figure*}[!tbp]
\centering
\centering
\subfigure[CIFAR-10 ($\text{ViT}_{\text{BASE}}$)]{
\includegraphics[width=0.25\textwidth]{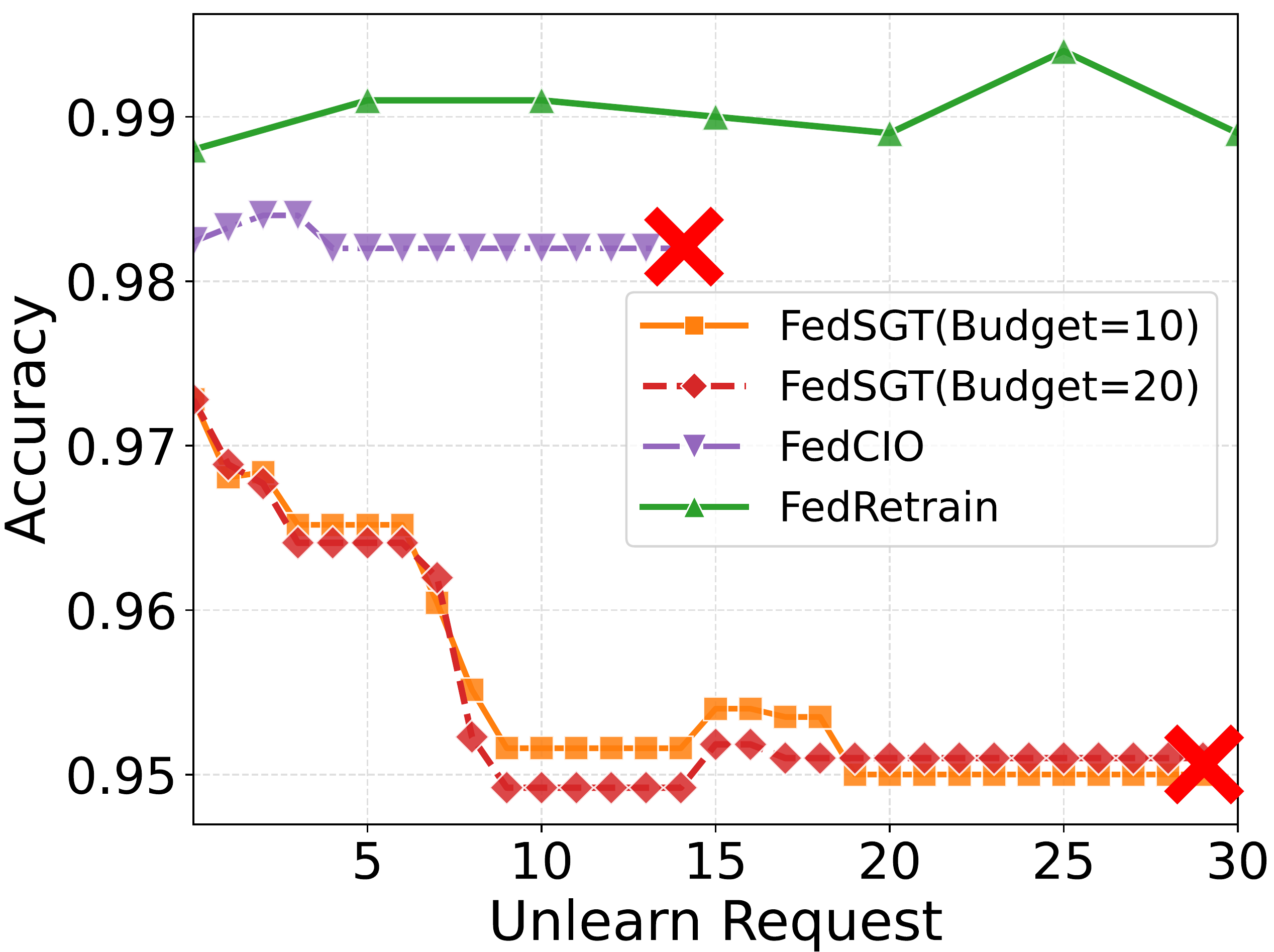}\label{fig:cifar10_iid_unlearn}}\hspace{0.5 em}
\subfigure[CIFAR-100 ($\text{ViT}_{\text{BASE}}$)]{
\includegraphics[width=0.25\textwidth]{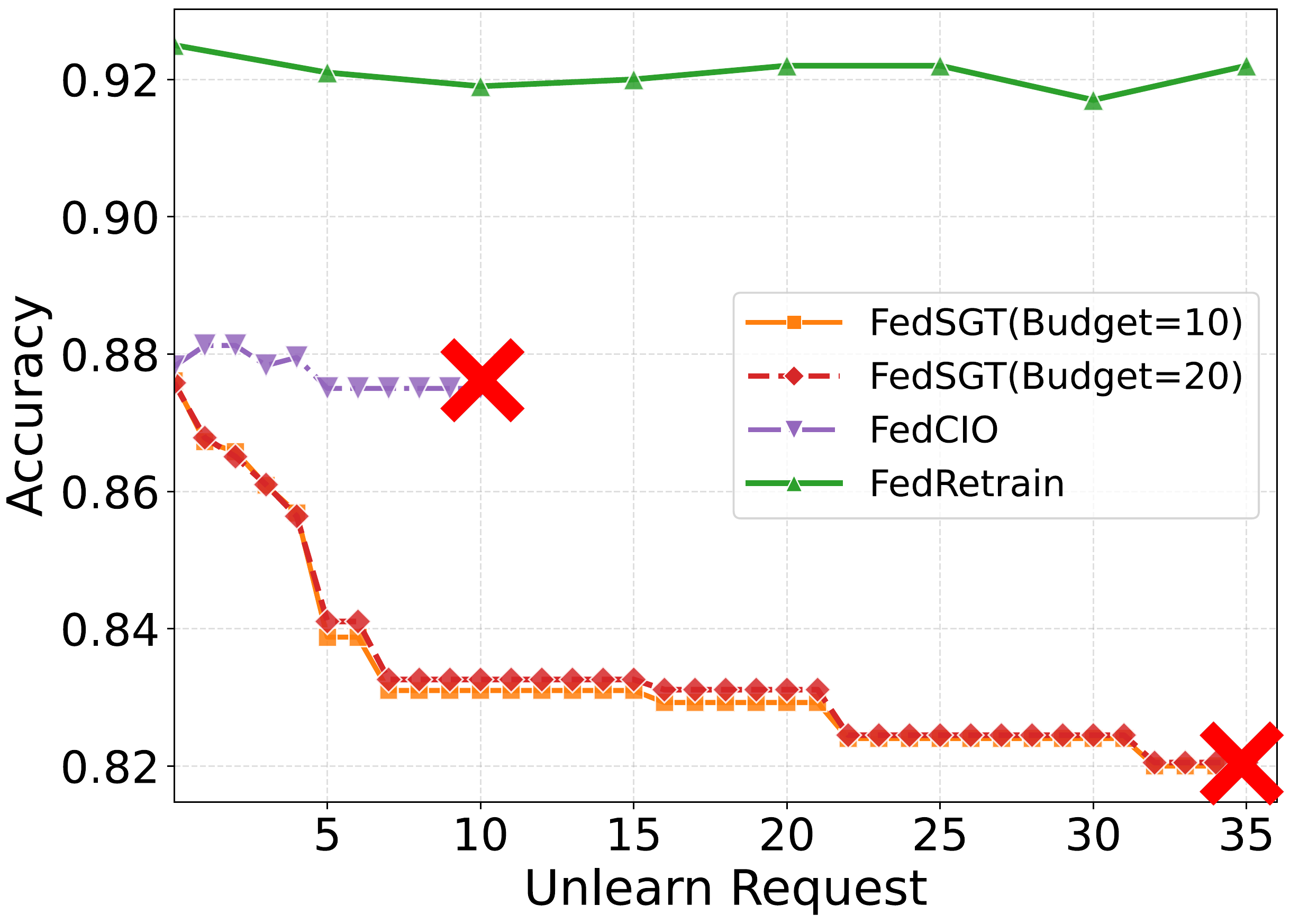}\label{fig:cifar100_iid_unlearn}}
\subfigure[SST-2 ($\text{RoBERT}_{\text{LARGE}}$)]{
\includegraphics[width=0.25\textwidth]{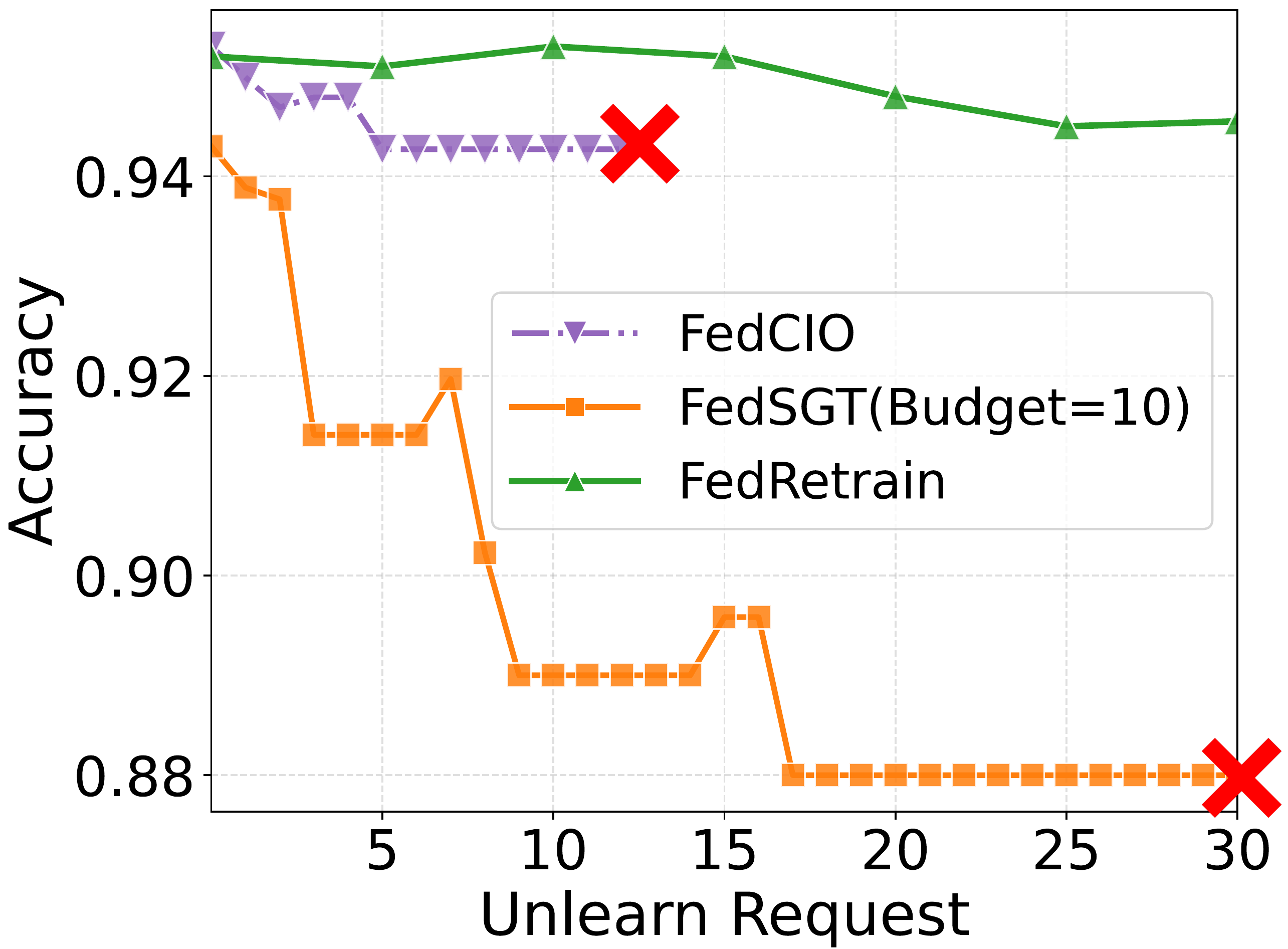}\label{fig:sst2_iid_unlearn}}\hspace{0.5 em}
\caption{Unlearning Performance on IID setting.}
\label{fig:unlearn_performance_iid}
\end{figure*}

\subsection{Training Performance}

We first evaluate the training performance of \method compared with the two baselines, FedAvg and FedCIO, under both vision and language tasks.  
Figure \autoref{fig:train_results_iid} reports the final test accuracy across all datasets, including CIFAR-10 and CIFAR-100 for vision, and four GLUE benchmarks (SST-2, QQP, QNLI, and MNLI) for natural language understanding.  

Overall, \method\ achieves comparable training performance than FedCIO and remains close to the standard FedAvg baseline, despite introducing the sequential group-based training mechanism.  
On CIFAR-10, \method\ almost matches the full fine-tuning accuracy of FedAvg, while maintaining a similar level of stability.  
On QQP task, \method achieves even slightly better performance than FedAvg and FedCIO, indicating that the introduction of lightweight PEFT modules and sequential updates does not compromise the model’s representation capacity.
These results confirm that \method\ retains the efficiency and effectiveness of standard federated fine-tuning. 

\mypara{Non-IID Setting}
We further evaluate all methods under Non-IID client data distributions to investigate the impact of data heterogeneity on training performance.  
As shown in Figure \autoref{fig:train_results_non_iid}, \method consistently achieves higher accuracy than both FedAvg and FedCIO across most tasks.  
This improvement stems from its uniform grouping strategy~\autoref{subsec:grouping}, which balances data representation across clients during sequential training.  
By forming more homogeneous groups and mitigating the bias introduced by skewed client distributions, 
\method effectively stabilizes the training process and alleviates performance degradation caused by Non-IID data.  
In contrast, FedCIO’s disjoint clustering amplifies inter-cluster heterogeneity, leading to less consistent optimization and inferior performance under Non-IID data.

\begin{figure*}[!tbp]
\centering
\centering
\subfigure[CIFAR-10 ($\text{ViT}_{\text{BASE}}$)]{
\includegraphics[width=0.25\textwidth]{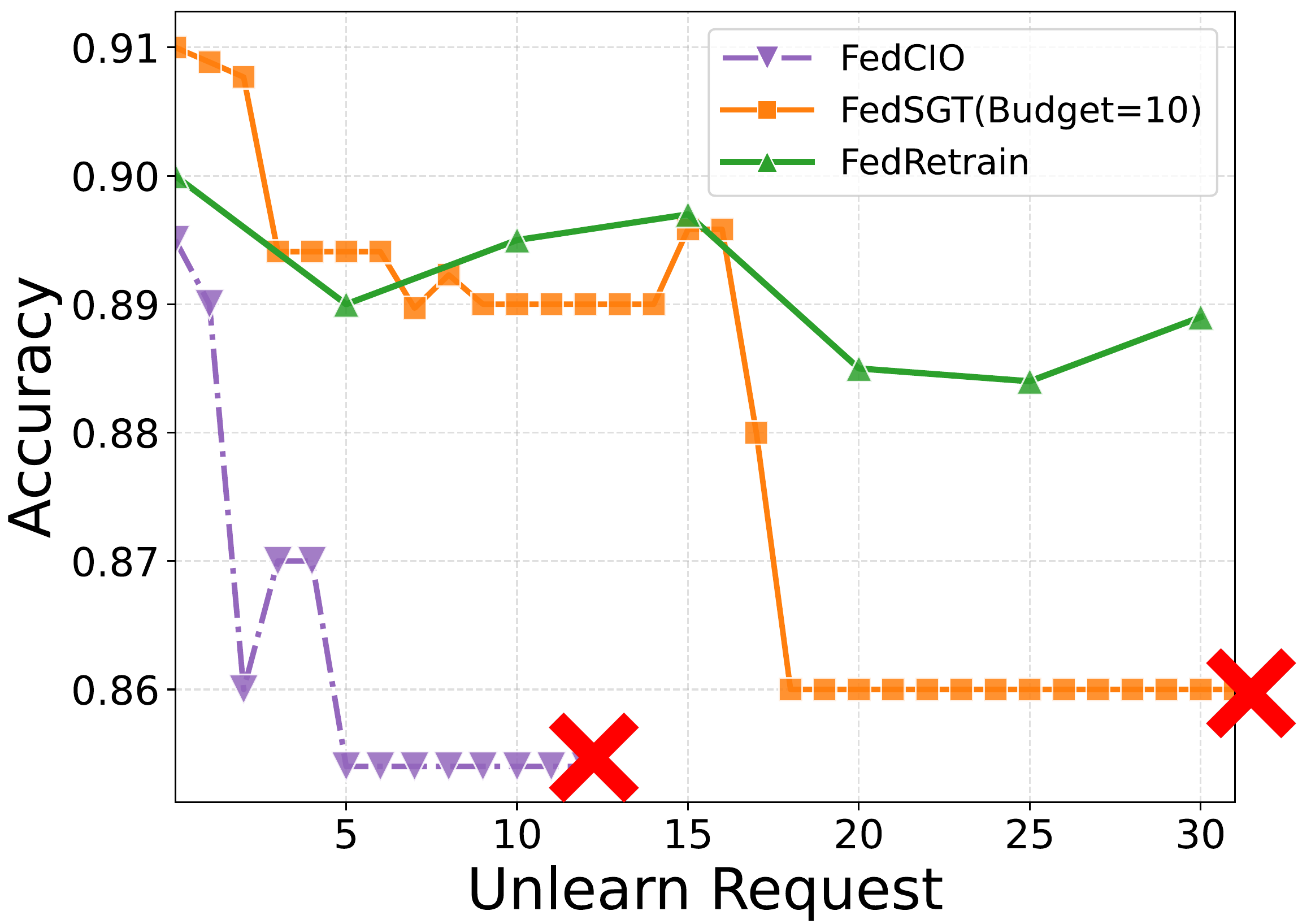}\label{fig:cifar10_noniid_unlearn}}\hspace{0.5 em}
\subfigure[CIFAR-100 ($\text{ViT}_{\text{BASE}}$)]{
\includegraphics[width=0.25\textwidth]{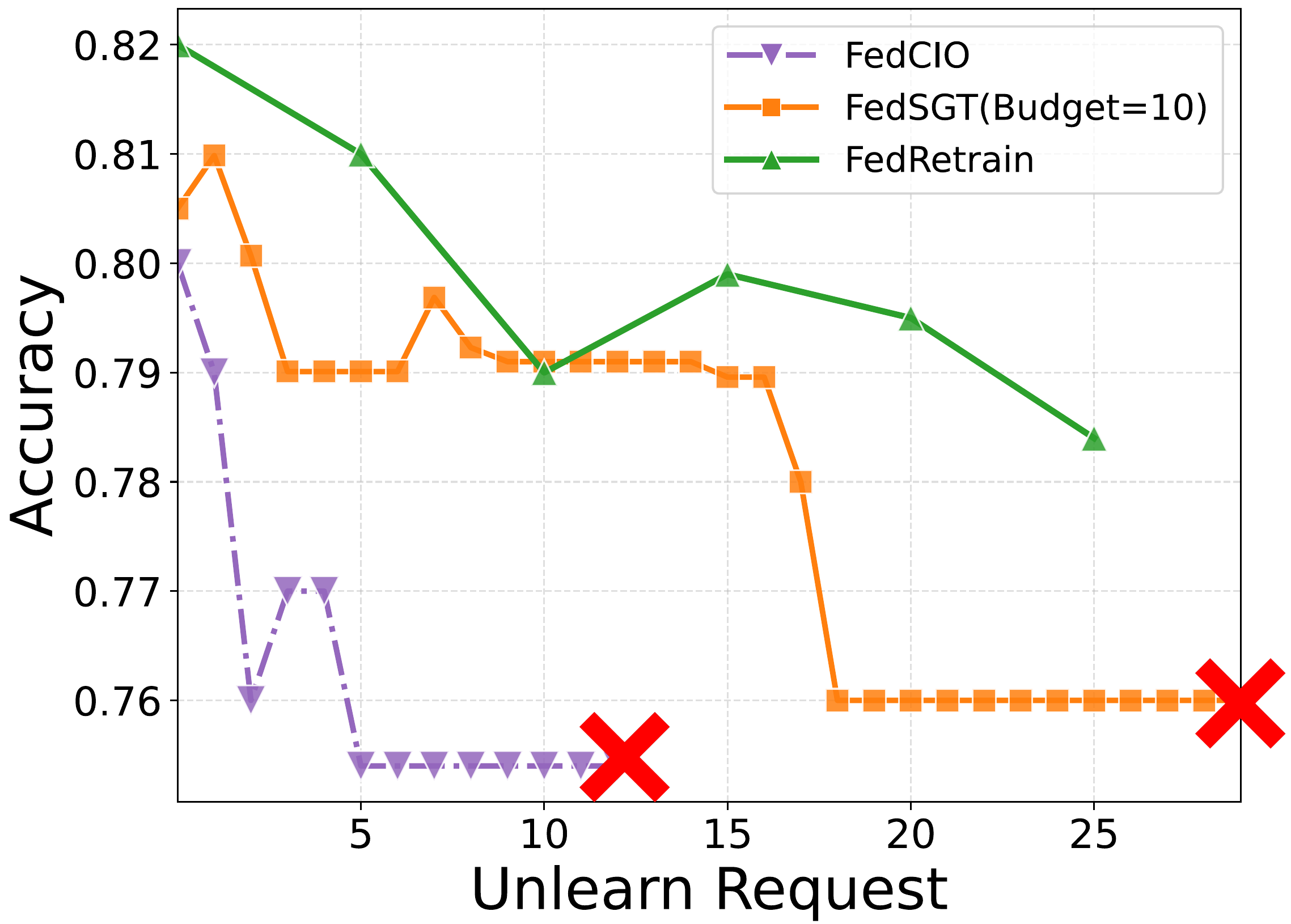}\label{fig:cifar100_noniid_unlearn}}
\subfigure[SST-2 ($\text{RoBERT}_{\text{LARGE}}$)]{
\includegraphics[width=0.25\textwidth]{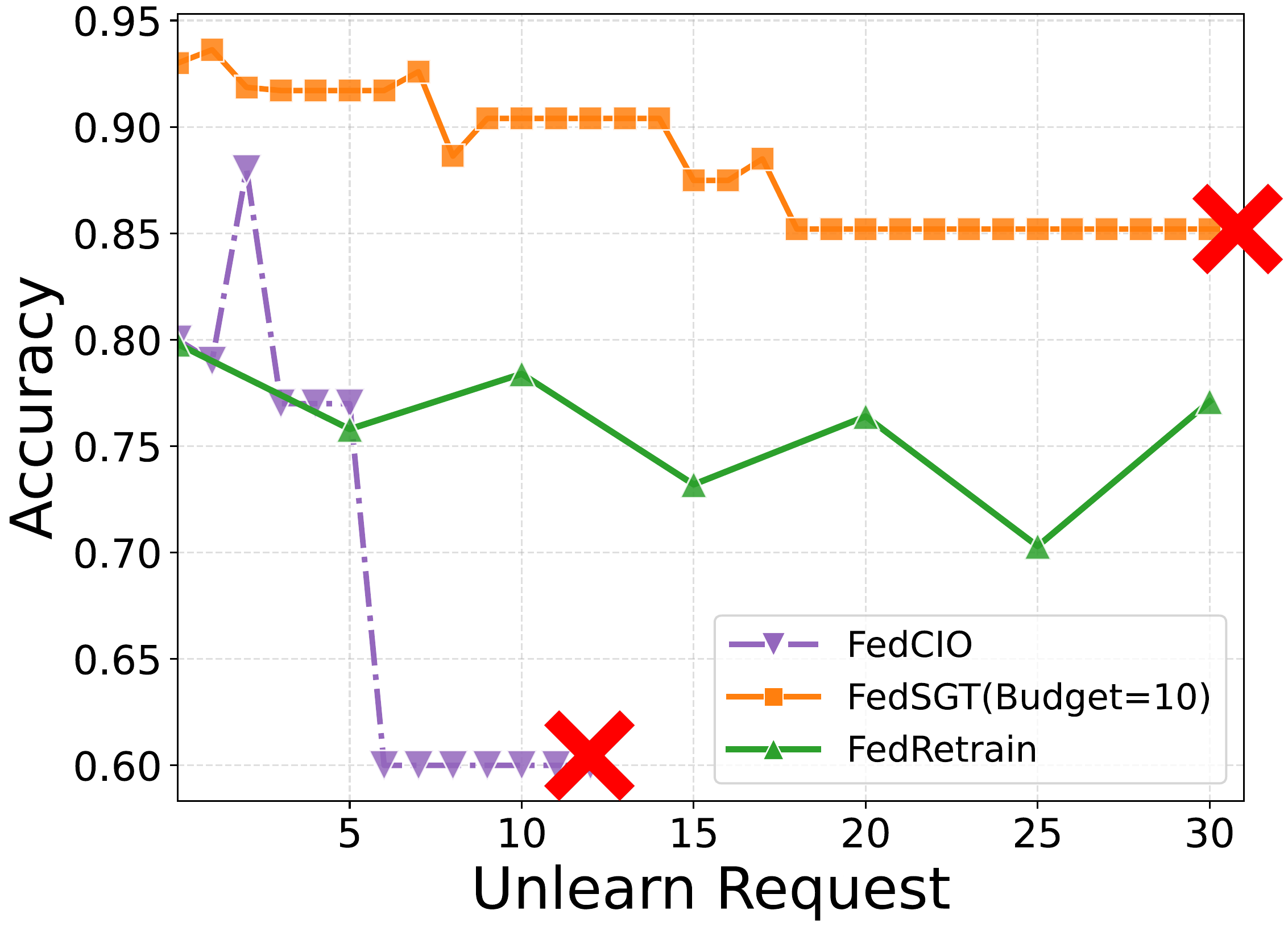}\label{fig:sst2_noniid_unlearn}}\hspace{0.5 em}
\caption{Unlearning Performance on Non-IID setting.}
\label{fig:unlearn_performance_noniid}
\end{figure*}

\begin{takeawaybox}
\textbf{Takeaway 1:} \textit{\method delivers comparable or even superior training performance to FedAvg and FedCIO under IID settings. Under Non-IID conditions, the uniform grouping strategy of \method effectively mitigates data heterogeneity, leading to more stable and better model performance.}
\end{takeawaybox}

\subsection{Unlearning Performance}

\mypara{Unlearning Request Setting}  
We assume that unlearning requests arrive independently, without any correlation among the deleted data samples.  
Each request targets a batch of 100 data records contained within a single client slice.

\autoref{fig:unlearn_performance_iid} reports the unlearning performance under IID data distributions.  
For clarity, we evaluate \textit{FedRetrain} every five unlearning requests due to its high retraining cost.  
Across all datasets, \method maintains service availability for substantially more deletion requests than FedCIO, which aligns with our theoretical deletion rate analysis in \autoref{subsec:theo_deletion_rate}.  
During early unlearning stages, FedCIO preserves slightly higher utility, as each cluster in the IID setting receives a nearly equal share of the global distribution.

\mypara{Non-IID Setting}
We further evaluate all methods under Non-IID setting to examine the impact of data heterogeneity on unlearning behavior.  
As shown in \autoref{fig:unlearn_performance_noniid}, \method consistently achieves higher model utility than FedCIO.  
FedCIO suffers from highly uneven cluster quality in the Non-IID regime: different clusters capture very different data distributions, causing their retrained or retained models to vary dramatically in accuracy.  
In contrast, \method remains stable because the proposed uniform random grouping strategy (\autoref{subsec:grouping}) distributes client slices across groups in a balanced manner, effectively mitigating the effects of Non-IID data during both training and unlearning.

\begin{takeawaybox}
\textbf{Takeaway 2:} \textit{
\method provides consistently longer service maintenance than FedCIO across both IID and Non-IID settings, supporting substantially more exact unlearning requests before failure. 
Moreover, under Non-IID distributions, \method achieves markedly higher model utility due to its uniform random grouping, which mitigates the adverse effects of data heterogeneity.}
\end{takeawaybox}

\begin{table}[!tbp]
\centering
\caption{
Training time (hours). \method\ uses $B=10$ sequences; parentheses show per-sequence cost.}
\label{tab:efficiency_time}
\begin{tabular}{l l 
                S[table-format=3.2]
                S[table-format=3.2]
                l}
\toprule
\textbf{Dataset} & \textbf{Model} & \textbf{FedAvg} & \textbf{FedCIO} & \textbf{\method}   \\
\midrule
CIFAR-10  & $\text{ViT}_{\text{BASE}}$      
          & 68.42 & 30.53 & 41.31\,(4.13) \\
CIFAR-100 & $\text{ViT}_{\text{BASE}}$      
          & 69.21 & 31.40 & 42.30\,(4.20) \\
SST-2     & $\text{RoBERT}_{\text{LARGE}}$ 
          & 53.40 & 20.30 & 46.20\,(4.62) \\
QQP       & $\text{RoBERT}_{\text{LARGE}}$ 
          & 182.44 & 112.35 & 181.0\,(18.10) \\
QNLI      & $\text{RoBERT}_{\text{LARGE}}$ 
          & 75.25 & 63.41 & 70.30\,(7.03)  \\
MNLI      & $\text{RoBERT}_{\text{LARGE}}$
          & 252.51 & 182.36 & 241.3\,(24.13) \\
\bottomrule
\end{tabular}
\end{table}

\begin{figure} [!tbp]
\centering  
\includegraphics[width=0.45\textwidth]{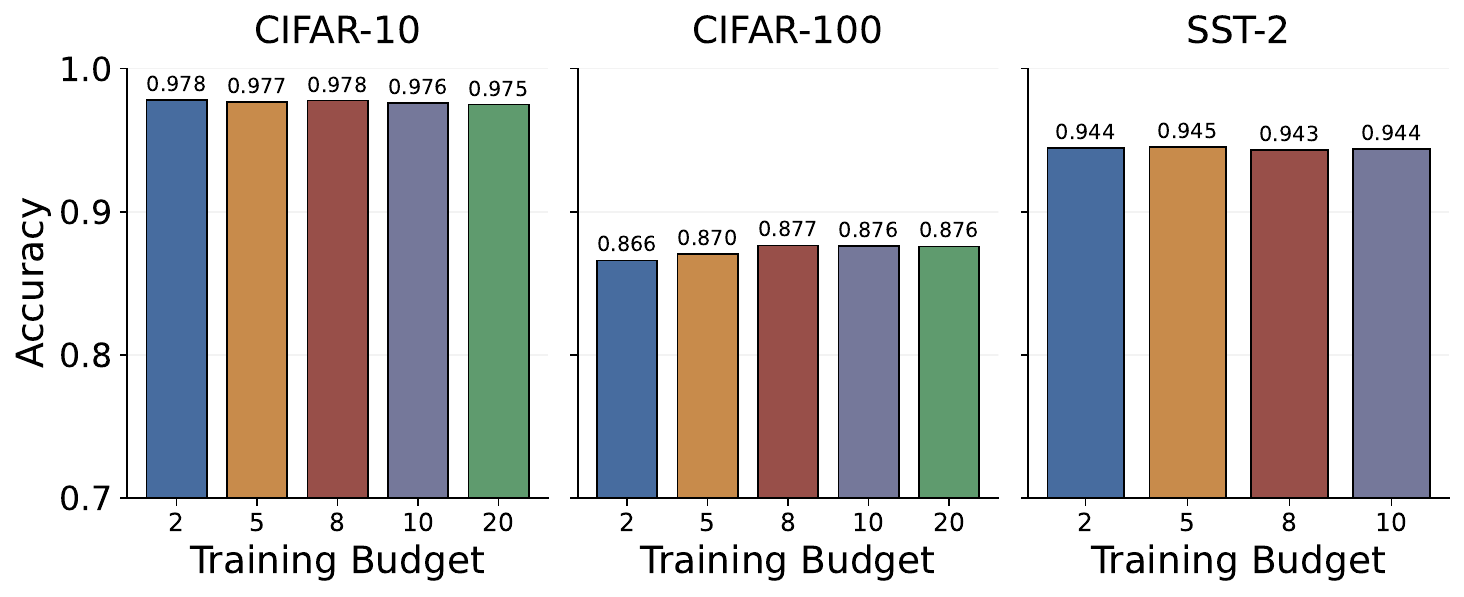}
\caption{Training Results with different training budgets.}
\label{fig:budget_analysis}
\end{figure}

\subsection{Training Efficiency}

\autoref{tab:efficiency_time} reports the end-to-end training time of all methods across
vision and language benchmarks. 
Overall, the total training cost of \method\ is comparable to FedAvg and FedCIO, despite the
larger absolute running time caused by the substantially heavier backbone models 
(e.g., $\text{ViT}_{\text{BASE}}$ and $\text{RoBERTa}_{\text{Large}}$).
This empirical observation is consistent with our complexity analysis in 
\autoref{subsec:overhead}: although \method\ trains $B{=}10$ sequences, each sequence
updates only a single lightweight PEFT module, and thus the per-sequence cost remains small 
(as reflected by the numbers in parentheses).

The deviations from the theoretical ratios primarily arise from practical system factors in our
standalone FL simulation environment, including client-side parallelism effects, heterogeneous
hardware throughput, and occasional communication stalls.
In particular, FedAvg and FedCIO benefit from fully parallel client updates within each round,
while \method's sequence-wise training introduces minor scheduling overhead. 
Nonetheless, the overall wall-clock time remains in the same order as FedAvg and FedCIO across all datasets,
confirming that \method\ achieves improved unlearning capabilities without incurring significant additional training cost.

\subsection{Ablation Study}

\subsubsection{Training Budget Analysis}
\autoref{fig:budget_analysis} reports the training accuracy of \method under different training budgets.
Overall, the performance remains stable across a wide range of budgets.
This behavior is expected: although larger budgets introduce more training sequences, each sequence corresponds to a different permutation of groups, and some permutations may lead to slightly stronger or weaker convergence.
As a result, increasing the budget does not necessarily yield monotonic improvement.
Instead, the accuracy variations stay within a small margin across all datasets, indicating that \method is inherently robust to the choice of training budget.
This also reflects the efficiency of our group-based training method: most useful information is already captured within a small number of sequences.

\begin{figure}[t]
\centering
\centering
\subfigure[CIFAR-10 ($\text{ViT}_{\text{BASE}}$)]{
\includegraphics[width=0.2\textwidth]{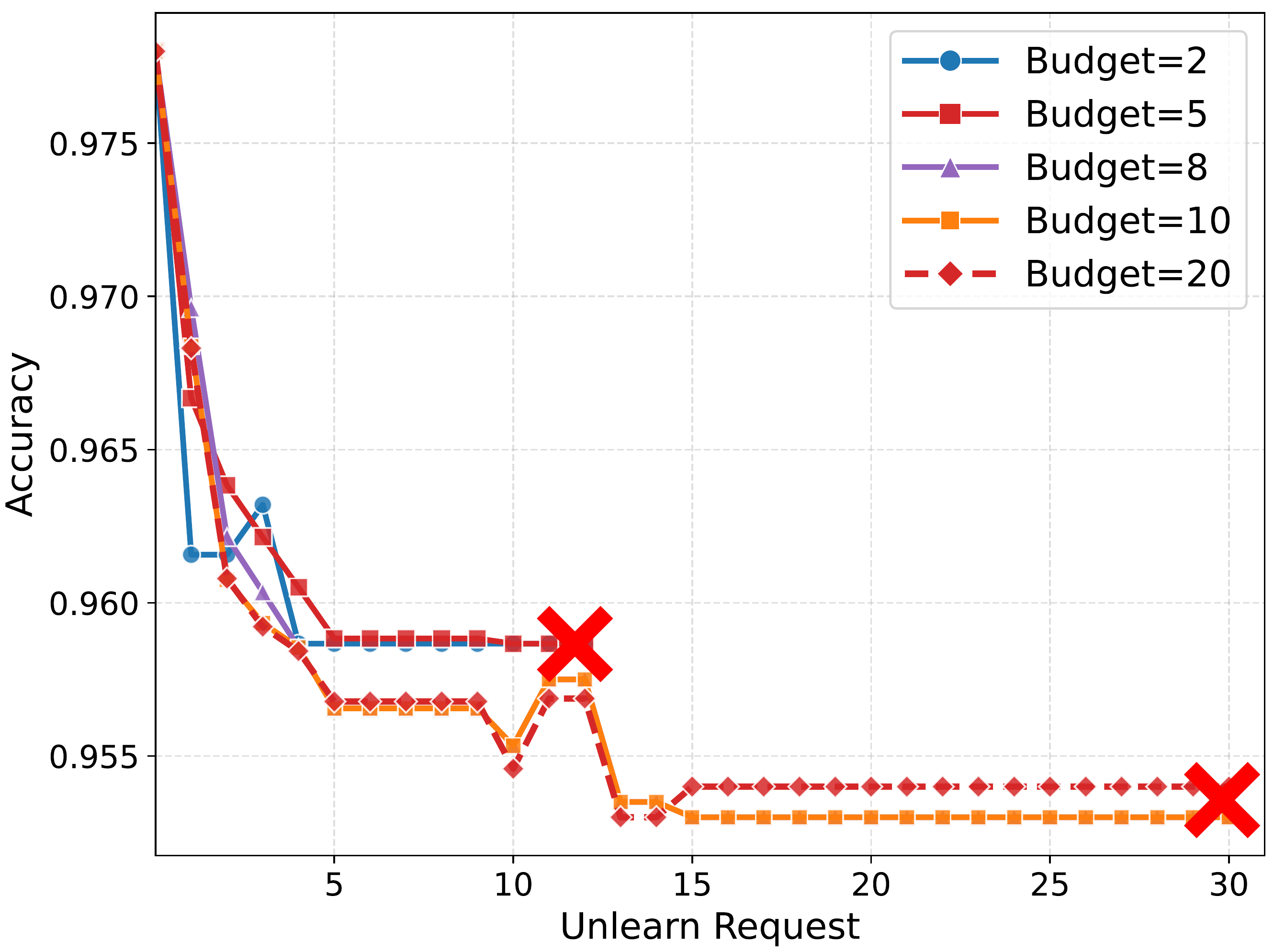}\label{fig:shards_r_c}}\hspace{0.5 em}
\subfigure[CIFAR-100 ($\text{ViT}_{\text{BASE}}$)]{
\includegraphics[width=0.2\textwidth]{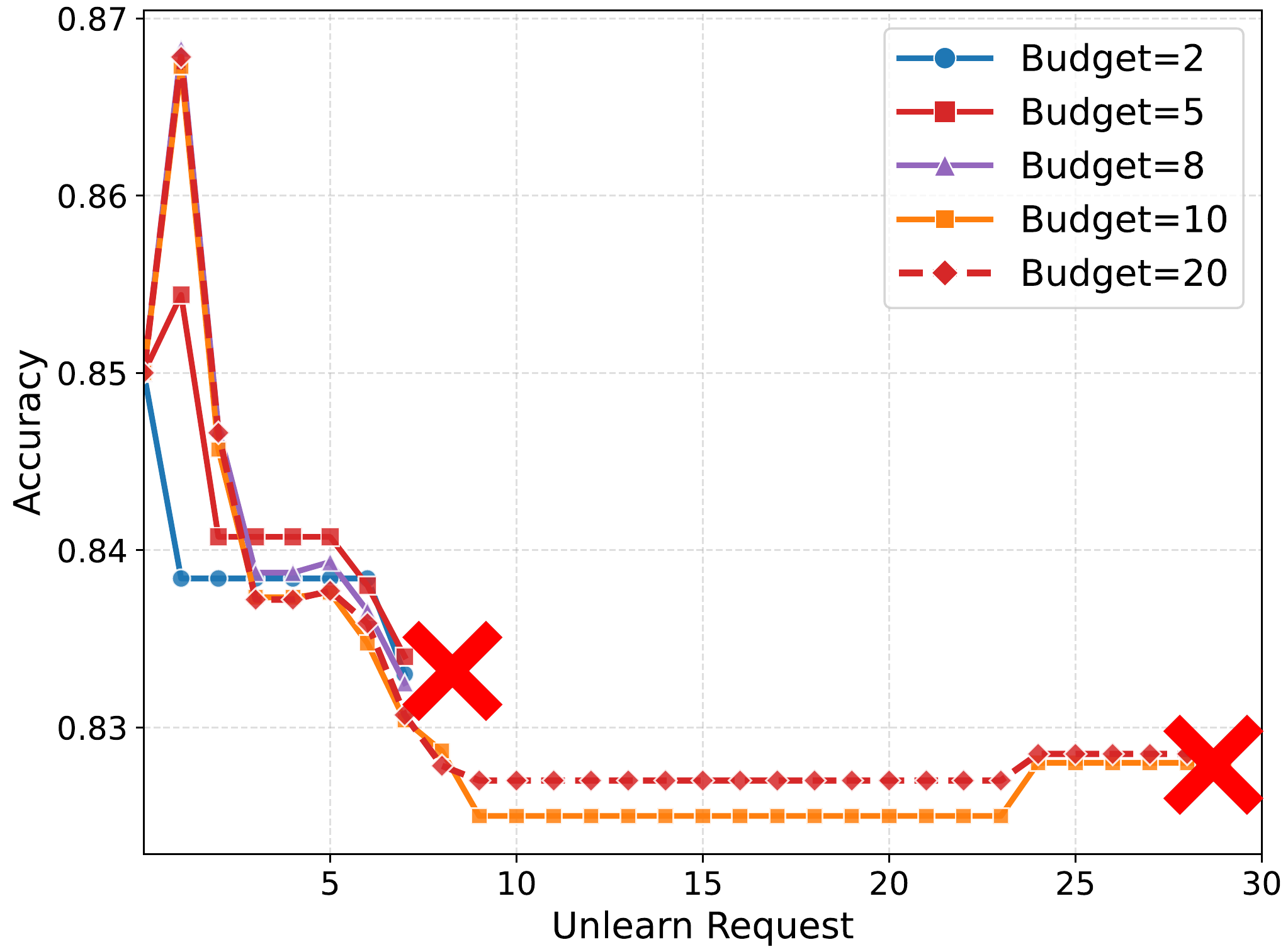}\label{fig:shards_r_s}}
\caption{\method's unlearning performance with different training budget constraints on IID setting.}
\label{fig:budget_unlearn}
\end{figure}

\mypara{Unlearning Performance}
\autoref{fig:budget_unlearn} illustrates the unlearning behavior of \method under different training budgets.
When the budget is smaller than the number of groups ($B < 10$), the system exhausts available sequences more quickly, leading to an earlier service termination.
As the budget increases, \method consistently supports a larger number of deletion requests, since more sequences remain valid after each unlearning event.

When the budget exceeds the group count ($B \ge 10$), the service maintenance time saturates and no longer improves with additional sequences.
This is expected: beyond $L$ sequences, many permutations become redundant, and the marginal gain in surviving deletions diminishes.

In terms of model utility during the unlearning process, smaller budgets may occasionally exhibit slightly better accuracy in the early stages due to randomness in sequence selection.
However, generally larger budgets yield more stable performance and significantly longer service availability.

\begin{takeawaybox}
\textbf{Takeaway 3:} \textit{
Increasing the training budget yields limited gains in training accuracy, but substantially improves unlearning robustness—larger budgets allow \method to sustain significantly longer service availability under repeated deletion requests.}
\end{takeawaybox}

\subsubsection{Client Slicing Number} 
\label{subsubsec:ablation_client_slice}

\begin{figure} [!tbp]
\centering  
\includegraphics[width=0.45\textwidth]{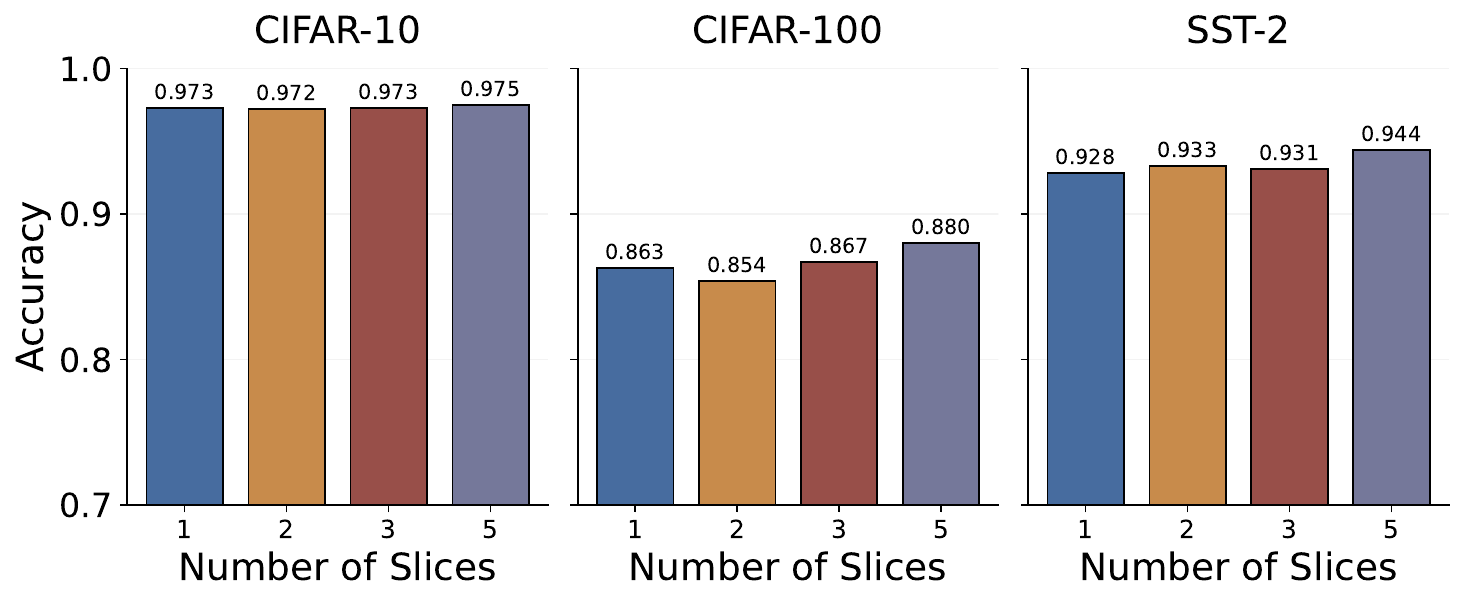}
\caption{Training Performance with various client slice numbers under IID setting}
\label{fig:slice_number_iid}
\end{figure}

\begin{figure} [!tbp]
\centering  
\includegraphics[width=0.45\textwidth]{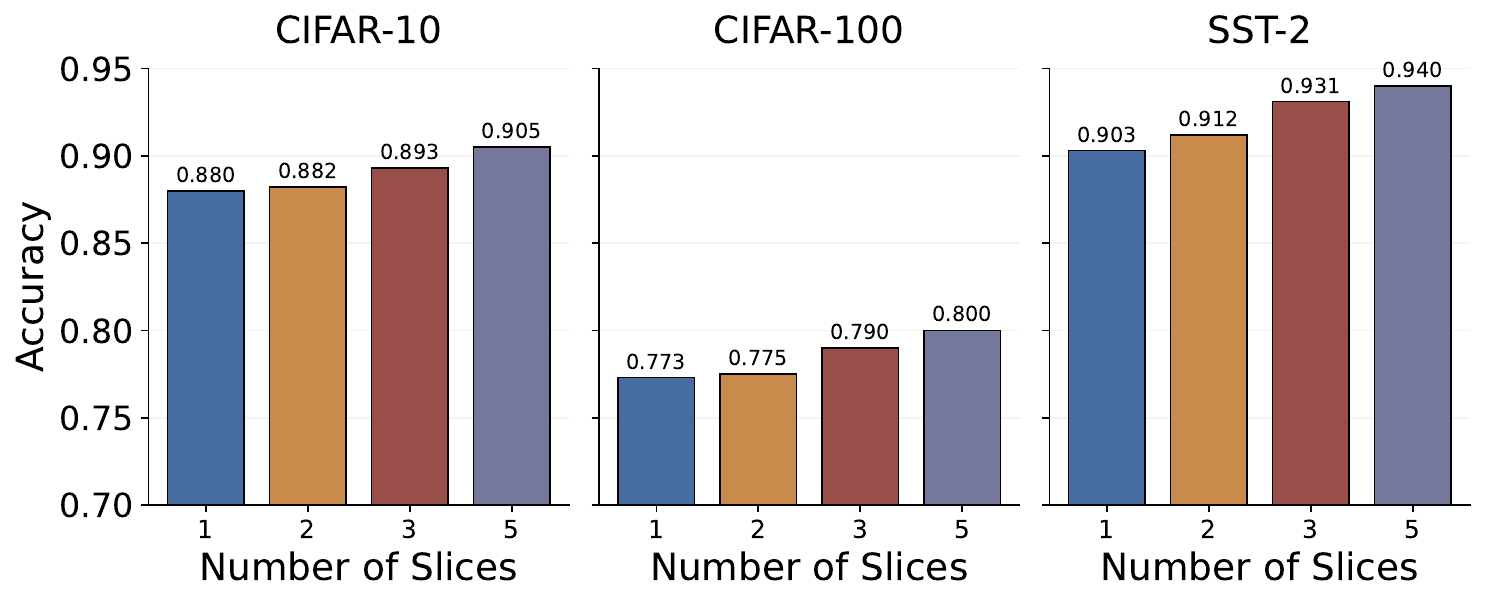}
\caption{Training Performance with various client slice numbers under Non-IID setting}
\label{fig:slice_number_noniid}
\end{figure}

\autoref{fig:slice_number_iid} and \autoref{fig:slice_number_noniid} illustrate the impact of different client slice numbers on the training performance of \method.
Under the IID setting (\autoref{fig:slice_number_iid}), varying the number of slices has only a minor effect on accuracy.
Since client data are already well-balanced across the population, additional slicing does not substantially alter the group distributions and thus leads to comparable performance across all slice counts.

In contrast, under the Non-IID setting (\autoref{fig:slice_number_noniid}), increasing the slice number leads to noticeably higher accuracy.
This improvement arises because finer slicing enables the uniform grouping strategy (\autoref{subsec:grouping}) to redistribute heterogeneous client data more evenly across groups.
As a result, each training phase encounters a more representative mixture of data, mitigating the adverse effects of data heterogeneity and improving overall model convergence.

\begin{takeawaybox}
\textbf{Takeaway 4:} \textit{
Client slicing matters more under Non-IID data. 
Under IID, different slice numbers yield similar performance. 
Under Non-IID, more slices substantially improve accuracy by balancing group composition. 
The tradeoff is higher communication participation as clients join more groups.}
\end{takeawaybox}

\begin{figure}[t]
\centering
\centering
\subfigure[CIFAR-10 ($\text{ViT}_{\text{BASE}}$)]{
\includegraphics[width=0.2\textwidth]{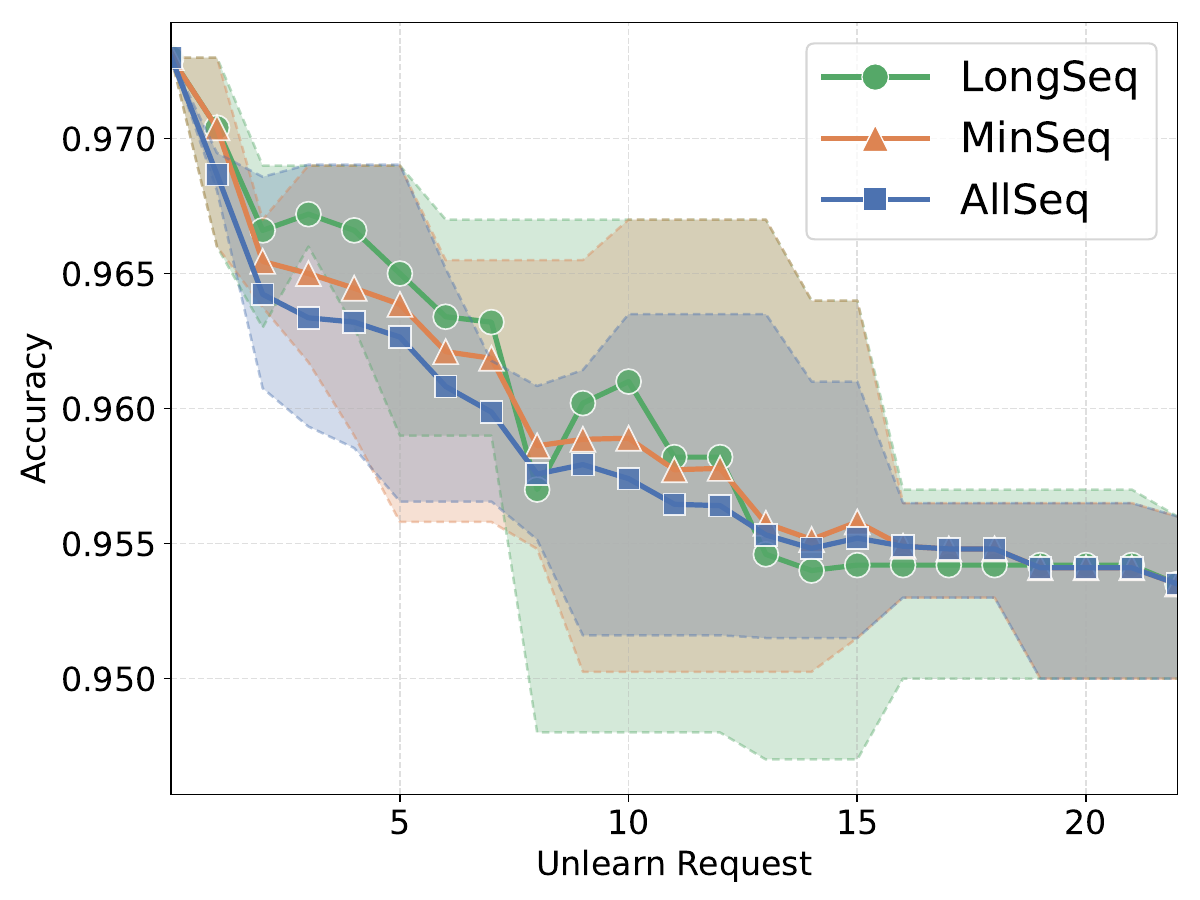}}\hspace{0.5 em}
\subfigure[CIFAR-100 ($\text{ViT}_{\text{BASE}}$)]{
\includegraphics[width=0.2\textwidth]{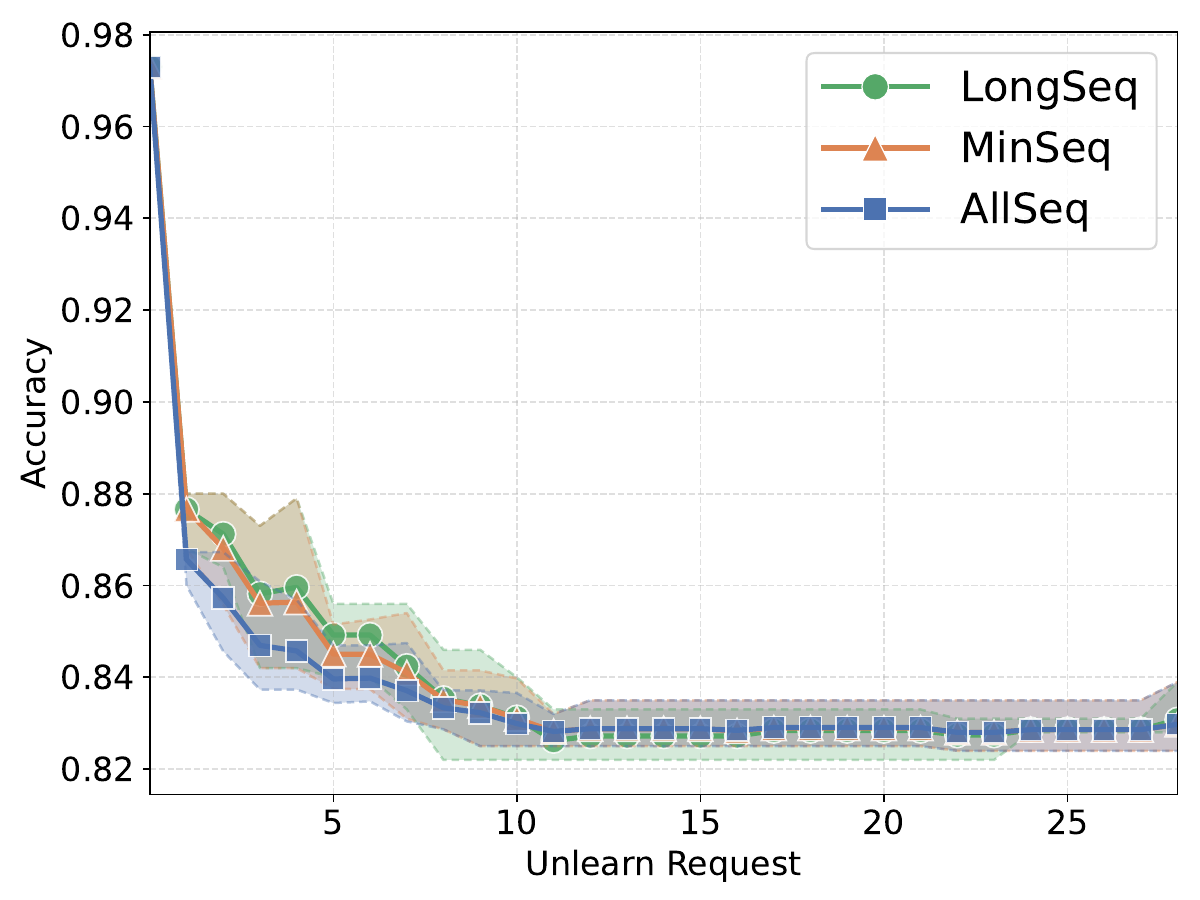}}
\caption{\method's unlearning performance with different sequence selection strategies.}
\label{fig:sequence_method_unlearn}
\end{figure}

\subsubsection{Sequence Selection Method}
We compare three sequence selection strategies—\textit{LongSeq}, \textit{MinSeq}, and \textit{AllSeq}—to understand how the choice of active sequence affects unlearning performance. 
As shown in \autoref{fig:sequence_method_unlearn}, all three strategies yield similar average performance across unlearning requests, indicating that the sequence-aware design of \method is generally robust to the selection rules.

However, the variability among strategies differs noticeably.  
\textit{LongSeq}, which always selects the longest remaining sequence, exhibits the largest variance: depending on how many valid modules remain in that sequence, it may perform unusually well or degrade more rapidly.  
In contrast, \textit{MinSeq} and \textit{AllSeq} provide more stable behavior, as they rely on either choosing the shortest surviving sequence or averaging over all valid sequences, both of which reduce variance during deletion progression.
We include further discussion in~\autoref{appendix:discussion}.

\begin{takeawaybox}
\textbf{Takeaway 5:} \textit{
All sequence-selection strategies achieve similar average unlearning performance. 
LongSeq shows higher variance because relying on a single sequence can lead to either very strong or very weak outcomes, whereas MinSeq and AllSeq offer more stability.}
\end{takeawaybox}

\section{Conclusion}
We presented \method, an exact federated unlearning framework that partitions client data into balanced groups and maps each group to lightweight PEFT modules across multiple training sequences. 
By isolating data groups at the module level, \method enables instant exact unlearning—deactivating modules tied to deletion requests without retraining—while keeping client storage and communication bounded. Theoretically, we analyze deletion rate, performance under accumulating deletions, and overhead; empirically, \method matches or exceeds baselines in accuracy and training efficiency and sustains longer service cycles under continual unlearning. Extensive ablations confirm robustness to data heterogeneity and hyper-parameters.

\bibliographystyle{ACM-Reference-Format}
\bibliography{sample}

\appendix
\section*{APPENDIX}

\section{Detailed Theoritical Analysis}

\subsection{Detailed Deletion Rate Derivation}
\label{sec:appendix_deletionrate}

\begin{proof}
Although \method differs from S3T in its training method, it adopts a similar sharding and slicing configuration. 
In \method, however, the number of \emph{shards} is fixed at $1$, and multiple slices from different clients jointly form a \emph{group}, which serves the same functional role as a slice in S3T. Each unlearning request randomly affects one cluster, following a similar mechanism as in S3T.

The deletion rate is the expected number of requests to delete all the topmost $B'=\min(L,B)$ slices. Using linearity of expectation, the deletion rate is 
\begin{align}
    \delta(\method)&=\mathbb{E}[t_1+t_2+\cdots+t_{B'}]\\
    &=\mathbb{E}[t_1]+\mathbb{E}[t_2]+\cdots+\mathbb{E}[t_{B'}]
    \label{eq:time-expectation}
\end{align}
where $E[t_i] = 1/p_i$. The denominator $p_i$ is the probability that the $i$-th slice is affected after $(i-1)$ slices are affected, which is given by

\begin{equation*}
     p_i\approx \frac{B'-(i-1)}{L}
\end{equation*}
The approximation in the above equation is based on the assumption that the number of unlearning requests is small enough that the group sizes remain unchanged. Substituting the above result into Equation~\eqref{eq:time-expectation}, we have:
\begin{align*}
    \delta(\method)
    &=\frac{L}{B'-0}+\frac{L}{B'-1}+\cdots\frac{L}{1}\\
    &=L\big(\frac{1}{1}+\frac{1}{2}+\cdots+\frac{1}{B'}\big)\\
    &\equiv L\cdot H_{B'}
\end{align*}

Similarly, deletion rate for FedCIO is the expected number of requests to affect all the $c$ clusters:
\begin{align}
   \delta(\text{FedCIO})&=\mathbb{E}[t_1+t_2+\cdots+t_{c}]\\
    &=\mathbb{E}[t_1]+\mathbb{E}[t_2]+\cdots+\mathbb{E}[t_{c}]
    \label{eq:time-expectation-FedCIO}
\end{align}
where $E[t_i] = 1/p_i$. Since there are totally $c$ clusters, the expression $p_i$ is given by:
\begin{equation*}
    p_i\approx \frac{c-(i-1)}{c}
\end{equation*}
Again, we assume that the number of unlearning requests is much smaller than the cluster size. Substituting the expression of $p_i$ into Equation~\eqref{eq:time-expectation-FedCIO}, we get:
\begin{align*}
\delta(\text{FedCIO})&=\frac{c}{c-0}+\frac{c}{c-1}+\cdots\frac{c}{1}\\
    &=c\big(\frac{1}{1}+\frac{1}{2}+\cdots+\frac{1}{c}\big)\\
    &\equiv c\cdot H_{c}
\end{align*}
\end{proof}

\subsection{Unlearning Performance Analysis}
\label{sec:appendix_performance}
\begin{proof}


To simulate the unlearning process, let variable $X_i$ denote the cluster id that the $i$th unlearning request affects, which is uniformly drawn from the integers from $0$ to $L-1$, where $i$ is from $1$ to $r$. 
Following the cyclic rotation algorithm in S3T, we want to find the number of remaining training samples in the sequence with the largest remaining clusters. Therefore, we define \textit{cyclic span ($U$)} as the minimum number of slices to be removed in a sequence using the cyclic rotation algorithm with $B\ge L$. It is mathematically written as $U = \min_j\{\max_i[(X_i+j) \% L] - \min_i[(X_i+j)\% L]\}$ where $j$ is an integer from $0$ to $L-1$. Now, we want to find the expectation of $T$ because the number of the largest remaining training samples is 
\begin{equation} \label{eq:R}
    R=\frac{|D|}{L}\cdot (L-T)
\end{equation}

A cleaner, equivalent way to write $T$ (and easier to reason about) is: Let $Z={z_1<z_2<\dots<z_m}$ be the distinct sampled values among $(X_1,\dots,X_r)$. Define the cyclic gaps
\begin{equation*}
    d_i=\begin{cases} z_{i+1}-z_i,& i=1,\dots,m-1\\
    z_1+L-z_m,& i=m \end{cases}
\end{equation*}
Then $T=L-\displaystyle\max_{1\le i\le m} d_i+1$.
The intuition is as follows: Rotating by $j$ is the same as choosing where to ``cut'' the circle into a line segment ($[0,L)$). The best cut is across the largest empty arc between consecutive occupied points; the span that needs to cover the points is the complement of that gap, i.e., $(L-\max d_i)+1$. 

Let $M$ be the number of distinct occupied positions (so $1\le M\le \min(r,L)$). Then 
\begin{equation}
    \Pr(M=m)=\binom{L}{m} \cdot m! \cdot S(r,m)/L^r
\label{eq:Pr(M=m)}
\end{equation}
, where $S(r,m)$ is the Stirling number of the second kind.
Conditional on $M=m$, the ordered gaps $(d_1,\dots,d_m)$ are uniformly distributed over all $m$-tuples of positive integers summing to $L$. Hence, for any $s\ge1$,

\begin{align}
\label{eq:Pr(max_d<=s|M=m)}
& \Pr\big(\max_i d_i \le s \big| M=m\big) \\
= & \frac{1}{\binom{L-1}{m-1}} \times \nonumber 
 \sum_{j=0}^{\left\lfloor\frac{L-m}{s}\right\rfloor}(-1)^j\binom{m}{j}\binom{L-1-j s}{m-1}
\end{align}

From this,
\begin{align}
    \mathbb{E}[U\mid M=m]&=L-\mathbb{E}[\max_i d_i\mid M=m]+1 \label{eq:E[T|M=m]}\\
    &=1+L-\sum_{t=1}^{L}\Pr\big(\max_i d_i\ge t\mid M=m\big)\label{eq:Pr(max d>=t|M=m}\\
    &=1+\sum_{s=1}^{L-1}\Pr\big(\max_i d_i\le s\mid M=m\big) \label{eq:Pr(max d<=s|M=m)}
\end{align}
and finally, by the definition of expectation,
\begin{equation}
\mathbb{E}[U]=\sum_{m=1}^{\min(r,L)} \Pr(M=m) \cdot \mathbb{E}[U\mid M=m].
\end{equation}

Applying linearity of expectation to Equation~\eqref{eq:R}, we have $\mathbb{E}[R]=\frac{|D|}{L}(L-\mathbb{E}[U])$.
\end{proof}

To clarify the above equations, we provide a detailed explanation below. To interpret Equation~\eqref{eq:Pr(M=m)}, consider the $r$ draws as placing $r$ labeled balls into $L$ labeled bins indexed by $(0, \dots, L-1)$. Since each draw is equally probable, there are $L^r$ possible assignments. The event $M = m$ corresponds to the case in which exactly $m$ bins are nonempty.
To compute the probability of this event, we first select which $m$ bins are occupied, which can be done in $\binom{L}{m}$ ways. We then distribute the $r$ labeled balls among these $m$ bins such that none remains empty. The number of surjective mappings from an $r$-element labeled set to an $m$-element labeled set is $m!S(r,m)$, where $S(r,m)$ denotes the Stirling number of the second kind. Specifically, $S(r,m)$ represents the number of ways to partition the $r$ labeled elements into $m$ nonempty unlabeled subsets, while the multiplicative factor $m!$ accounts for assigning these subsets to the $m$ labeled bins.

The denominator of Equation~\eqref{eq:Pr(max_d<=s|M=m)} is the total number of possible positive $m$-tuples that sum up to $L$. It is equivalent to select $m-1$ bars out of $L-1$ possible positions to divide $L$ objects into $m$ partitions. Next, we count those with $\max d_i\leq s$. Write $z_i=d_i-1$. Then $z_i\ge 0$, $\sum z_i=L-m$, and $\max d_i\le s \iff z_i\le s-1$ for all $i$.
Let ($A_i$) be the event ($z_i\ge s$). For a fixed set ($J$) of size ($j$), substitute ($z_i'=z_i-s$) for ($i\in J$). Then
\begin{equation*}
    \sum_{i\notin J} z_i + \sum_{i\in J} z_i' = (L-m)-js
\end{equation*}
which has $\binom{(L-m)-js+m-1}{m-1}=\binom{L-1-js}{m-1}$ solutions if $(L-m)-js\ge 0$, thus $j$ take values from $0$ to $\lfloor (L-m)/s\rfloor$. By the inclusion–exclusion principle,
\begin{equation*}
    \#\{\max d_i\le s\}=\sum_{j=0}^{\lfloor (L-m)/s\rfloor} (-1)^j \binom{m}{j}\binom{L-1-js}{m-1}.
\end{equation*}

Equation~\eqref{eq:E[T|M=m]} comes from applying linearity of expectation to its definition. From Equation~\eqref{eq:E[T|M=m]} to Equation~\eqref{eq:Pr(max d>=t|M=m}, we leverage the fact that $d_i$ is an non-negative integer-valued random variable and applied the tail-sum formula $\mathbb{E}[X]=\sum^\infty_{k=1}\Pr(X\ge k)$, where $X$ is a non-negative random integer.
To see why from Equation~\eqref{eq:Pr(max d>=t|M=m} to Equation~\eqref{eq:Pr(max d<=s|M=m)}, let $F_m(s)\equiv\Pr\big(\max_i d_i\le s\mid M=m\big)$, whose expression is Equation~\eqref{eq:Pr(max_d<=s|M=m)} for $s \geq 1$ and $0$ for $s=0$, i.e., $F_m(0)=0$. Then 
\begin{align*}
   \Pr(\max_i d_i\ge t\mid M=m)
   &=1-\Pr(\max_i d_i\le t-1\mid M=m)\\
   &=1-F_m(t-1) 
\end{align*}
Substitute it in to Equation~\eqref{eq:Pr(max d>=t|M=m},
\begin{align*}
    \mathbb{E}[U\mid M=m]
    &=1+L-\sum_{t=1}^{L}\Pr\big(\max_i d_i\ge t\mid M=m\big)\\
    &=1+L-\sum_{t=1}^{L}\bigl[1-F_m(t-1)\bigr]\\
    &=1+\sum_{t=1}^{L}F_m(t-1)
\end{align*}
Renaming $s=t-1$ gives a clean form 
\begin{equation*}
    \mathbb{E}[U\mid M=m]=1+\sum_{s=0}^{L-1} F_m(s)=1+\sum_{s=1}^{L-1} F_m(s)
\end{equation*}
, where the last equality is because $F_m(0)=0$ as defined earlier. By substituting the definition of $F_m(s)$, we get Equation~\eqref{eq:Pr(max d<=s|M=m)}.

\subsection{Detailed Communication Cost Derivation}
\label{app:comm-derivation}
In this section, we derive the full expression of the expected communication cost for a client under uniform random grouping.
Consider a client partitioned into $S$ slices, and recall that all $NS$ slices 
are uniformly assigned to $L$ groups following~\autoref{alg:random-grouping}. 
Let $K$ denote the number of distinct groups that contain at least one slice of this client. 
The total communication cost is determined by~$K$.

\subsubsection*{(1) Communication cost conditioned on $K=k$}
Suppose the client belongs to exactly $k$ groups. 
Across the $L$ cyclic training sequences, let $X_1,\dots,X_k$ denote the positions 
of these groups in a given sequence, and let $V=\min_i X_i$. 
Once the first such group appears (position $V$), the client participates in all 
subsequent rounds, yielding a per-sequence cost of $L-V+1$. 
Since the $k$ positions form a uniform subset of $\{1,\dots,L\}$, the minimum 
order statistic satisfies $\mathbb{E}[V]=(L+1)/(k+1)$, and therefore
\[
\mathbb{E}[C_{\text{seq}}\mid K=k]=(L+1)\frac{k}{k+1}.
\]
Since \method executes all $L$ cyclic sequences,
\begin{equation}
\mathbb{E}\!\left[C_{\text{total}}(L)\mid K=k\right]
= L(L+1)\frac{k}{k+1}.
\label{eq:cost-K-app}
\end{equation}

\subsubsection*{(2) Distribution of $K$ under random grouping}
Each of the $S$ labeled slices independently selects one of the $L$ groups, 
yielding $L^S$ possible assignments. 
To obtain exactly $k$ occupied groups, we first choose which $k$ groups are used 
($\binom{L}{k}$ choices). 
Then we assign the $S$ slices onto these $k$ labeled groups such that each group 
receives at least one slice. 
The number of such onto assignments is $k!\,{S \brace k}$, where 
${S \brace k}$ is the Stirling number~\cite{graham1994concrete} of the second kind. 
Therefore,
\begin{equation}
\Pr(K=k)
=\frac{\binom{L}{k}\,k!\,{S\brace k}}{L^S},
\qquad k=1,\dots,\min\{S,L\}.
\label{eq:K-dist-app}
\end{equation}

\subsubsection*{(3) Final closed-form expectation}
Combining~\eqref{eq:cost-K-app} and~\eqref{eq:K-dist-app}, the unconditional 
expected communication cost is
\begin{align}
\mathbb{E}[C_{\text{total}}(L,S)]
&= \sum_{k=1}^{\min\{S,L\}}
\Pr(K=k)\,L(L+1)\frac{k}{k+1} \\
&= L(L+1)\sum_{k=1}^{\min\{S,L\}}
\frac{\binom{L}{k}\,k!\,{S\brace k}}{L^S}\frac{k}{k+1}.
\label{eq:final-cost-app}
\end{align}

Equation~\eqref{eq:final-cost-app} gives the exact expected communication overhead 
of a client as a function of $S$, the slicing granularity of \method.

\section{Performance-Overhead Trade-off under Heterogeneous FL Devices}
\label{appendix:analysis_edge}

In federated settings, edge devices can be highly heterogeneous in their computation and communication capabilities.
To better accommodate this heterogeneity, \method allows using different slicing granularities $S$ for different clients.
Intuitively, a larger $S$ provides finer-grained data partitioning and typically improves model performance (as observed in~\autoref{subsubsec:ablation_client_slice}), but it also increases the client-side communication overhead.
Hence, $S$ serves as a practical parameter to balance model quality and device affordability.

\mypara{Communication overhead increases with \(S\)}
The slicing granularity \(S\) affects the client communication overhead through the number of occupied groups \(K\).
As \(S\) increases, the client's slices are more likely to spread across more groups, which increases the expected number of occupied groups:
\[
\mathbb{E}[K]
= L\left(1-\left(1-\frac1L\right)^S\right),
\]
which is monotone increasing in \(S\).
Meanwhile, conditioned on \(K=k\), the expected communication cost is
\[
\mathbb{E}[C_{\text{total}}(L)\mid K=k]
= L(L+1)\frac{k}{k+1},
\]
which is strictly increasing in \(k\).
Therefore, the expected communication cost is monotonically non-decreasing in \(S\), increasing from
\[
\mathbb{E}[C_{\text{total}}(L,1)] = \frac{L(L+1)}{2}
\]
toward the saturation limit
\[
\lim_{S\to\infty}\mathbb{E}[C_{\text{total}}(L,S)] = L^2.
\]

\mypara{Training cost is relatively insensitive to $S$}
Although a larger \(S\) makes a client appear in more groups (thus increasing communication/activation rounds), it does not necessarily increase local training computation.
The reason is that the client's total local data size remains fixed; slicing only changes how the same data are distributed across groups.
Under uniform random grouping and cyclic ordering, each slice has the same expected reuse count across phases, so the expected training cost mainly depends on the total local data size, rather than the slicing granularity \(S\).

\mypara{Practical implication}
This property enables a flexible deployment strategy for heterogeneous edge devices:
clients with stronger communication budgets can use a larger $S$ to obtain better model performance, while communication-constrained devices can use a smaller $S$ to reduce communication overhead, with only limited impact on local training cost.

\section{Ablation Study About Other PEFT}\label{appendix:other_peft_ablation}

In addition to LoRA, we consider two representative PEFT strategies that are widely used in NLP models: \emph{Prefix Tuning} and \emph{adapters}. 
\textbf{Prefix Tuning} optimizes a small set of continuous \emph{virtual tokens} (prefixes) that are prepended to the key/value states (or input embeddings) of each Transformer layer, while keeping the backbone frozen. This approach steers model behavior by modifying the attention context with a compact number of trainable parameters controlled by the prefix length.
\textbf{Adapters} generally insert lightweight task-specific modules into Transformer blocks while keeping the backbone frozen. 
In our implementation, we instantiate this adapter-style PEFT baseline using \textbf{IA3} (Infused Adapter by Inhibiting and Amplifying Inner Activations~\cite{liu2022few}), which adapts the model by learning multiplicative scaling vectors on internal activations rather than inserting conventional bottleneck layers. 
For clarity and consistency, we refer to this baseline as IA3 in the remainder of this section.

\mypara{Setup}
To fairly compare different PEFT choices in \method, we keep the federated training pipeline and data grouping identical across methods, and only change the PEFT-specific modules and their associated optimization hyperparameters.
All experiments are conducted on SST-2 using the same backbone model ($\text{RoBERT}_{\text{LARGE}}$).

We use Prefix Tuning with 20 virtual tokens and a prefix hidden size of 512.
We optimize the prefix parameters with a learning rate of \(5\times10^{-3}\), while freezing the backbone model parameters.
IA3 introduces lightweight multiplicative scaling parameters on internal activations while keeping the backbone frozen.
We optimize IA3 parameters with a learning rate of \(1\times10^{-3}\).

\begin{figure} [!tbp]
\centering  
\includegraphics[width=0.38\textwidth]{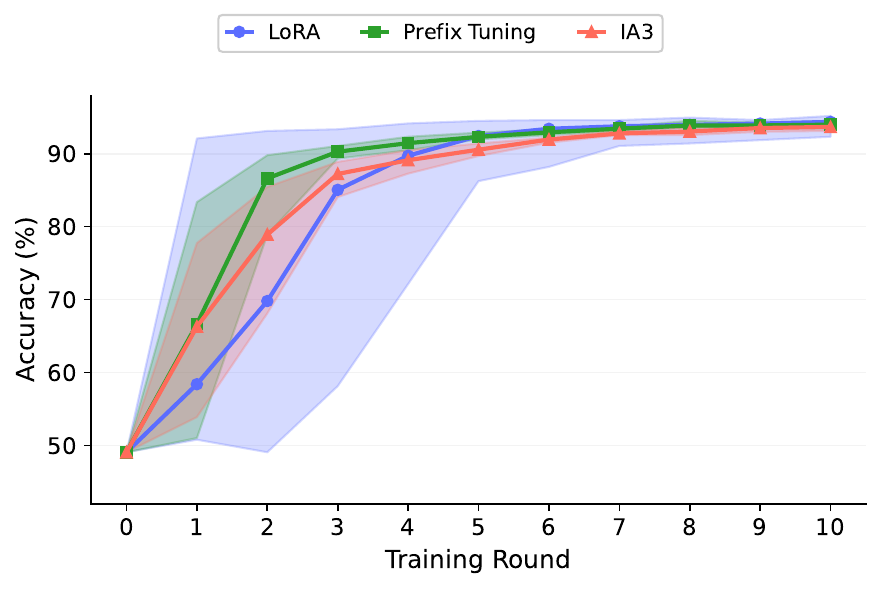}
\caption{Learning curve of different PEFT in \method}
\label{fig:peft_ablation_results}
\end{figure}

\mypara{Learning Performance}
\autoref{fig:peft_ablation_results} shows the learning curves of \method with three PEFT choices (LoRA, Prefix Tuning, and IA3) under different sequence orders.
Overall, all three methods achieve very similar final utility on SST-2, indicating that \method\ is compatible with multiple PEFT backbones and does not rely on a specific adaptation module.
In particular, LoRA achieves the best final accuracy ($0.938$), while Prefix Tuning and IA3 obtain $0.935$ and $0.932$, respectively.
Although LoRA performs slightly better at performance, the margin is small, suggesting comparable end-task performance across the three PEFT strategies.

In terms of training dynamics, Prefix Tuning and IA3 exhibit more stable trajectories across rounds (and across different orders), with smoother improvement and smaller fluctuation during early-to-mid training stages.
By contrast, LoRA shows larger variance in the training process, especially in the earlier rounds, but catches up quickly and eventually converges to the best final accuracy.
These results suggest a trade-off between \emph{training stability} and \emph{peak performance}: Prefix Tuning and IA3 provide more stable optimization behavior, whereas LoRA offers a slightly stronger final performance in our setting.

\begin{figure} [!tbp]
\centering  
\includegraphics[width=0.38\textwidth]{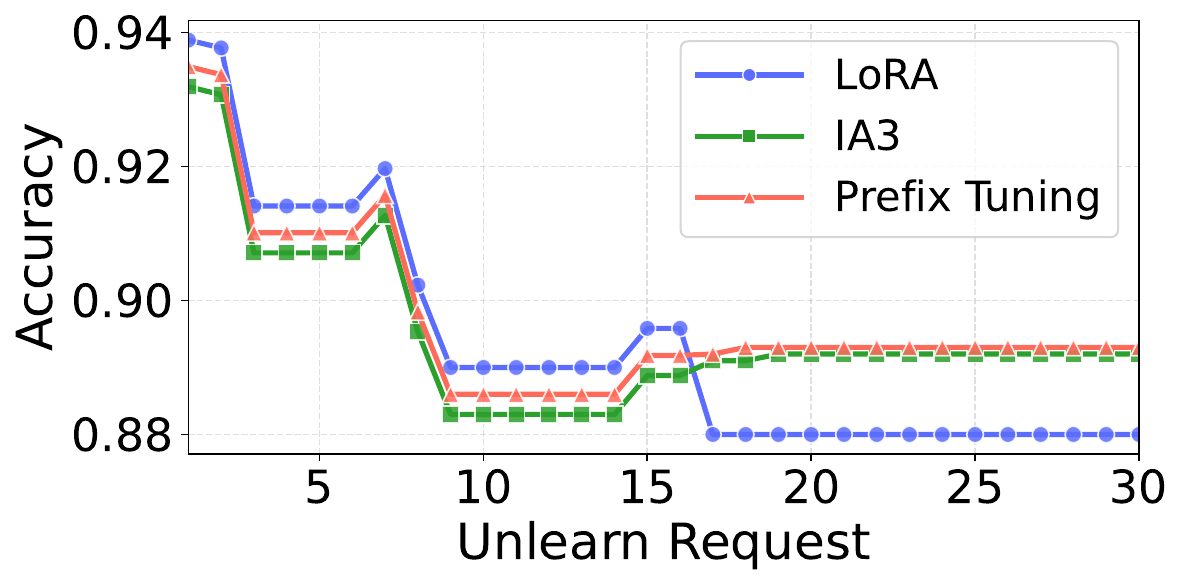}
\caption{Unlearniong curve of different PEFT in \method}
\label{fig:peft_unlearning_ablation}
\end{figure}

\mypara{Unlearning Performance}
\autoref{fig:peft_unlearning_ablation} compares the unlearning-time utility of \method with different PEFT choices under a sequence of unlearning requests.
Overall, the three curves exhibit highly similar trend patterns.
This is expected because all PEFT variants follow the same unlearning pipeline in \method: for each unlearning request, they select available checkpoints using the same sequence selection method as described in \autoref{subsec:unlearning}.
As a result, the major turning points and stage-wise transitions are aligned across methods.

In terms of utility during unlearning, the three PEFT methods remain close throughout the process, and their performance may alternately surpass one another under different requests.
This indicates that \method is broadly compatible with different PEFT methods in the unlearning stage, and no single PEFT choice dominates at every request step.
Meanwhile, in the later stage (when only one group of data remains available), Prefix Tuning and IA3 achieve slightly better utility than LoRA.
The reason is that Prefix Tuning and IA3 provide better adaptation capacity in this low-data regime, leading to stronger service utility after repeated unlearning updates.

\mypara{Efficiency}
\autoref{tab:ablation_time_sst2} reports the training time of \method\ on SST-2 with different PEFT choices.
Overall, the three methods have comparable training cost under the same federated pipeline, indicating that replacing the PEFT module does not substantially change the end-to-end runtime of \method.
Among them, LoRA is the fastest (4.62 hours per sequence), IA3 (Adapters) is slightly slower (5.38 hours per sequence), and Prefix Tuning takes the longest time (5.57 hours per sequence).

Although IA3 has a smaller trainable parameter set and may appear computationally lighter in principle, the observed wall-clock time is affected by more than parameter count alone.
In our federated implementation, the end-to-end runtime also includes framework-level overheads (e.g., PEFT module wrapping, optimizer/kernel efficiency, communication and synchronization, checkpoint saving, and evaluation), and these overheads can dominate or offset the theoretical savings of a lighter PEFT parameterization.
Similarly, Prefix Tuning introduces additional prefix-related operations in attention, which can increase practical runtime despite its parameter efficiency.
Therefore, our results suggest that wall-clock efficiency is implementation- and system-dependent, and LoRA provides the best practical efficiency among the three PEFT choices while all methods remain in a similar efficiency range.

\mypara{Summary}
Overall, the results indicate that \method\ is compatible with different PEFT choices, including LoRA, Prefix Tuning, and IA3.
Across these PEFT methods, the differences in learning performance, unlearning performance, and efficiency are relatively small in our SST-2 experiments.
This suggests that the effectiveness of \method does not rely on a specific PEFT instantiation.
Among them, LoRA shows a marginal advantage in final accuracy and runtime.
LoRA may also be a favorable default choice in practice due to its broad adoption, good architectural compatibility, and mature tooling support.

\newcolumntype{Y}{>{\centering\arraybackslash}X}

\begin{table}[!tbp]
\centering
\footnotesize
\renewcommand{\arraystretch}{1.1}
\caption{Training time (hours) of \method\ on SST-2. \method\ uses $B=10$ sequences; parentheses show per-sequence cost.}
\label{tab:ablation_time_sst2}
\begin{tabularx}{\columnwidth}{l Y Y Y}
\toprule
\textbf{Dataset} & \textbf{LoRA} & \textbf{Prefix Tuning} & \textbf{Adapters} \\
\midrule
SST-2 & 46.20 (4.62) & 55.74 (5.57) & 53.79 (5.38) \\
\bottomrule
\end{tabularx}
\end{table}

\section{Ablation Study About Grouping Method}
\label{appendix:group}

\begin{algorithm}[!tbp]
\caption{Barycenter-Aware Centering (BAC) Grouping Method}
\label{alg:BAC grouping}
\textbf{Input:} Client set $\mathcal{C}=\{c_1,\dots,c_N\}$; pretrained encoder $E(\cdot)$; per-client slice counts $\{D_i\}_{i=1}^{N}$; number of base clusters $K$; number of groups $G$\\
\textbf{Output:} Groups $\mathcal{G}=\{G_1,\dots,G_G\}$\\
\textbf{Procedures:}
\begin{algorithmic}[1]
    \STATE \textbf{(Client-side slicing and embedding)}
    \FOR{\textbf{each} client $c_i \in \mathcal{C}$}
        \STATE Partition local data into slices $\{S_i^1,\dots,S_i^{D_i}\}$
        \FOR{$s=1$ \textbf{to} $D_i$}
            \STATE $e_i^s \leftarrow \frac{1}{|S_i^s|}\sum_{x\in S_i^s} E(x)$ \cmt{slice embedding}
        \ENDFOR
        \STATE Send $\{e_i^1,\dots,e_i^{D_i}\}$ to server \cmt{no raw data sharing}
    \ENDFOR

    \STATE \textbf{(Server-side equal-size semantic base clustering)}
    \STATE $E_{\mathrm{all}} \leftarrow \{e_i^s \mid i\in[1..N],\, s\in[1..D_i]\}$
    \STATE $M \leftarrow \sum_{i=1}^{N} D_i$ \cmt{total number of slices}
    \STATE $g \leftarrow \frac{1}{M}\sum_{i=1}^{N}\sum_{s=1}^{D_i} e_i^s$ \cmt{global centroid}
    \STATE Run balanced clustering on $E_{\mathrm{all}}$ to obtain $\mathcal{B}=\{B_1,\dots,B_K\}$ with $|B_l|=G$ for all $l$ \cmt{equal-size base clusters}

    \STATE \textbf{(Barycenter matching across base clusters)}
    \FOR{$j=1$ \textbf{to} $G$}
        \STATE $T_j^{(0)} \leftarrow \mathbf{0}$; $G_j \leftarrow \varnothing$ \cmt{initialize group sum and group set}
    \ENDFOR

    \FOR{$l=1$ \textbf{to} $K$}
        \STATE Order items in $B_l$ by a fixed deterministic key, and write $B_l=\{b_{l,1},\dots,b_{l,G}\}$
        \FOR{$j=1$ \textbf{to} $G$}
            \FOR{$u=1$ \textbf{to} $G$}
                \STATE $C_{j,u}^{(l)} \leftarrow \left\| T_j^{(l-1)} + e(b_{l,u}) - l\,g \right\|_2^2$ \cmt{stage-$l$ target sum}
            \ENDFOR
        \ENDFOR
        \STATE $\pi_l^\star \leftarrow \arg\min_{\pi_l \in \mathcal{S}_G} \sum_{j=1}^{G} C_{j,\pi_l(j)}^{(l)}$
        \STATE Solve the above one-to-one assignment using the Hungarian algorithm with deterministic tie-breaking
        \FOR{$j=1$ \textbf{to} $G$}
            \STATE $G_j \leftarrow G_j \cup \{b_{l,\pi_l^\star(j)}\}$
            \STATE $T_j^{(l)} \leftarrow T_j^{(l-1)} + e(b_{l,\pi_l^\star(j)})$
        \ENDFOR
    \ENDFOR
\end{algorithmic}
\textbf{Return:} Group set $\mathcal{G}=\{G_1,\dots,G_G\}$
\end{algorithm}

To study the impact of different grouping methods, we further design a \textbf{Barycenter-Aware Centering (BAC)} grouping method.
Different from conventional clustering methods that mainly emphasize \emph{intra-cluster similarity}, our goal here is different: we expect \emph{each group} to preserve as much \emph{global information} as possible.
This design is particularly desirable for the unlearning stage, because as groups are progressively removed due to unlearning requests, the remaining groups can still maintain a relatively high utility if each group is individually representative of the overall data distribution.

\begin{figure}[!tbp]
\centering
\centering
\subfigure[Random vs BAC grouping method's performance]{
\includegraphics[width=0.22\textwidth]{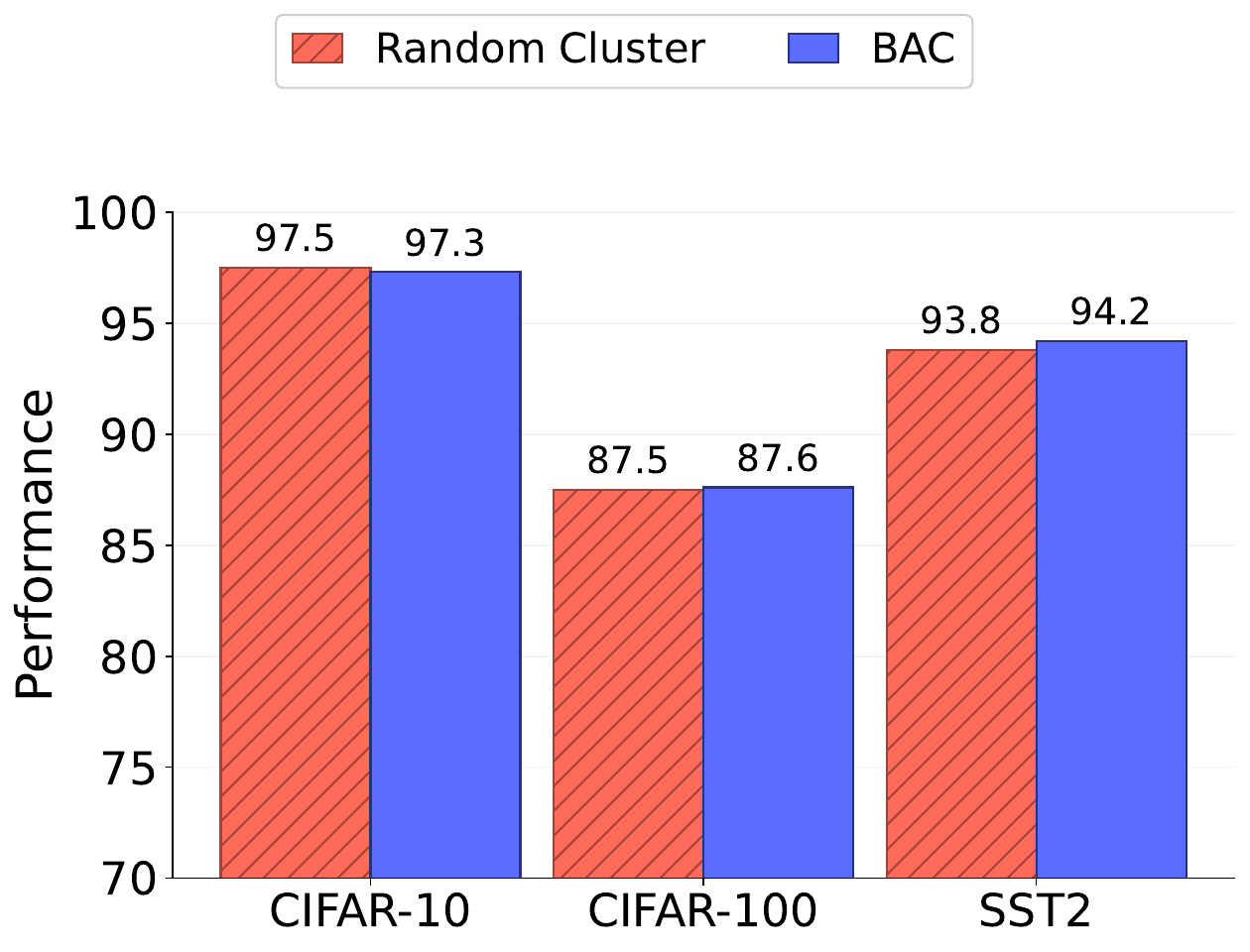}\label{fig:shards_r_c}}\hspace{0.5 em}
\subfigure[Random vs BAC grouping method's unlearning performance]{
\includegraphics[width=0.22\textwidth]{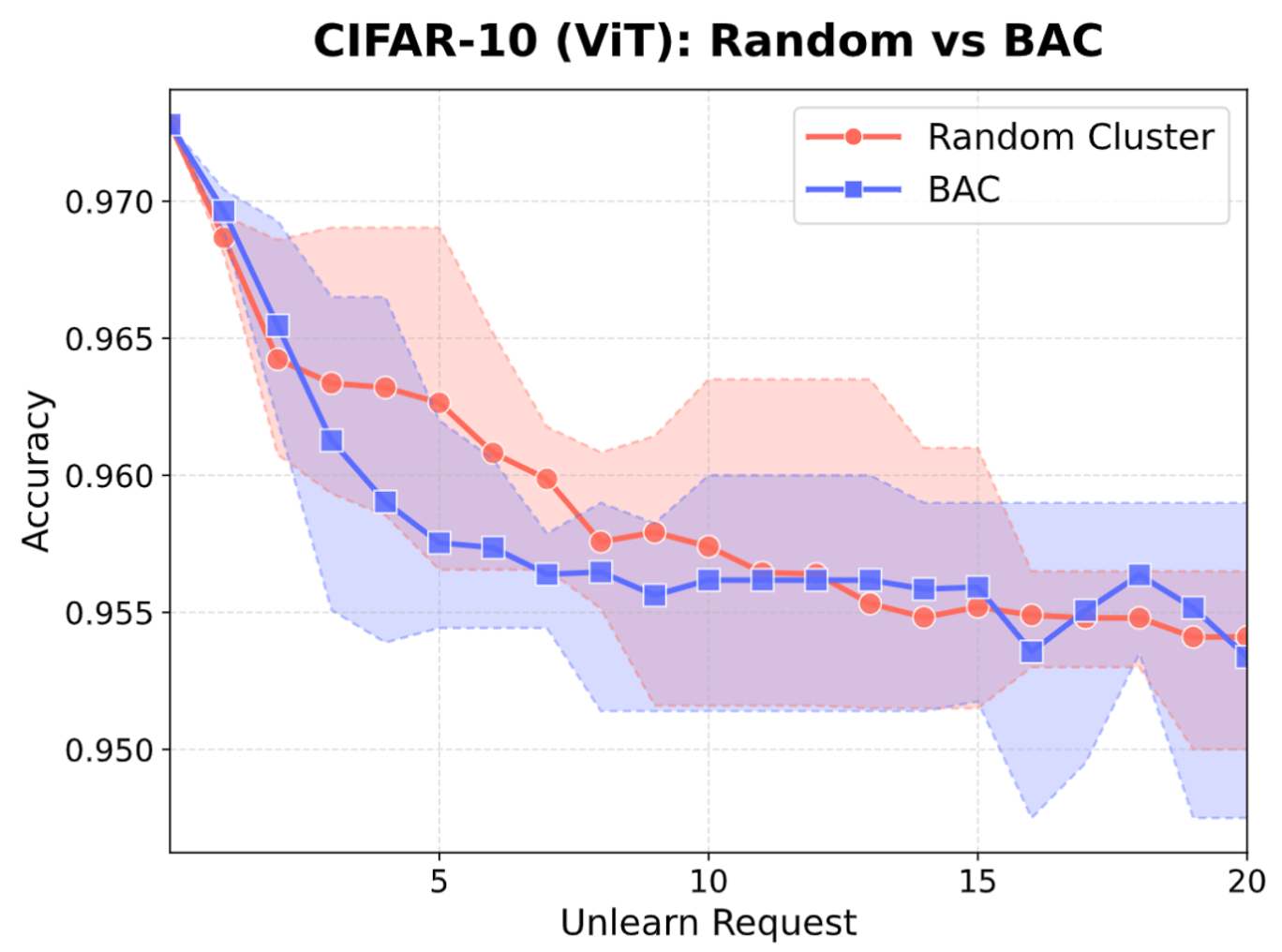}\label{fig:shards_r_s}}
\caption{\method's unlearning performance with different training budget constraints.}
\label{fig:bac_ablation}
\end{figure}

The detailed procedure is shown in \autoref{alg:BAC grouping}.
We first compute a slice-level embedding for each client slice using a pretrained encoder, and then aggregate all slice embeddings to obtain the \emph{global centroid} (i.e., the barycenter of all slices).
Instead of directly using standard clustering as the final grouping result, BAC first constructs \emph{equal-size semantic base clusters} over all slice embeddings, and then performs a deterministic matching process across these base clusters to form the final groups.
In particular, BAC assigns one slice from each base cluster to each group, while minimizing the deviation between each group's cumulative embedding sum and the stage-wise target centered at the global centroid.
As a result, the centroid of each final group is encouraged to be close to the global centroid, making each group more globally representative.

Intuitively, BAC can be viewed as a \emph{centroid-balanced grouping} strategy: it aims to preserve semantic diversity by drawing slices from different base clusters, while encouraging global representativeness through barycenter-aware matching.
By design, this property is expected to be beneficial in our setting, where the model needs to sustain service quality under repeated unlearning requests and progressive loss of group-level data.

\mypara{Effect of BAC Grouping}
We compare the proposed BAC grouping method with random grouping in terms of both learning performance and unlearning performance.
As shown in~\autoref{fig:bac_ablation}(a), BAC and random grouping achieve very similar learning performance across datasets (CIFAR-10, CIFAR-100, and SST-2), with only marginal differences in final utility.
This suggests that, under our current settings, replacing random grouping with BAC does not substantially change the standard training performance of \method.

We further compare their behavior during unlearning in~\autoref{fig:bac_ablation}(b).
Overall, BAC and random grouping exhibit similar utility trajectories, and BAC does not show a consistently large advantage in this experiment.
A possible explanation is that random grouping already produces groups that are sufficiently representative of the global data distribution in our setting (e.g., due to the slicing and shuffling procedure), which reduces the room for further improvement from barycenter-aware grouping.

From an efficiency perspective, BAC incurs additional preprocessing overhead in the grouping stage.
In CIFAR-10 experiments, the grouping time of BAC is \textbf{102} seconds, compared with \textbf{8} seconds for random grouping.
Therefore, although BAC is motivated by improving group-level representativeness, its empirical benefit is limited in our current experiments while introducing extra grouping cost.

Overall, these results suggest that random grouping is already sufficient in our current setting, offering a better trade-off between efficiency and effectiveness.

\begin{table*}[!tbp]
\centering
\caption{The hyperparameters used in experiments for \method, FedCIO and FedAvg across different benchmarks and Transformer model variants.}
\label{tab:hyperparams}
\begin{tabularx}{\columnwidth}{@{} l >{\raggedright\arraybackslash}X c c c @{}}
\toprule
\textbf{Dataset} & \textbf{Model} & \textbf{Rank ($r$)} & \textbf{Alpha ($\alpha$)} & \textbf{Learning Rate} \\
\midrule
\multicolumn{5}{l}{\textit{\textbf{Vision Tasks}}} \\
CIFAR-10   & $\text{ViT}_{\text{BASE}}$ & 16 & 16 & 2e-3 \\
CIFAR-100  & $\text{ViT}_{\text{BASE}}$ & 16 & 16 & 2e-3 \\
\midrule
\multicolumn{5}{l}{\textit{\textbf{Language Tasks (GLUE)}}} \\
SST-2      & $\text{RoBERT}_{\text{LARGE}}$ & 16  & 32 & 1e-4 \\
QQP        & $\text{RoBERT}_{\text{LARGE}}$ & 16  & 32 & 1e-4 \\
QNLI       & $\text{RoBERT}_{\text{LARGE}}$ & 16  & 32 & 1e-4 \\
MNLI       & $\text{RoBERT}_{\text{LARGE}}$ & 16  & 32 & 1e-4 \\
\bottomrule
\end{tabularx}
\end{table*}

\section{Related Work}
\label{appendix:related_work}

\mypara{Approximate Federated Unlearning}
Many FU methods use approximate strategies that reduce a client's influence without guaranteeing exact removal.  
A major line of work is gradient-ascent(GA)–based unlearning~\cite{EWCSGA, FUsurvery2}, which increases the forgotten client's loss, but direct GA often severely harms the utility of retained clients~\cite{FUPGA}.  
To address this, EWCSGA~\cite{EWCSGA} adds regularization to limit utility drop.  
Another family computes update directions orthogonal to the input subspace~\cite{saha2021gradient, SFU}, which works in centralized settings but is unsuitable for FL due to privacy leakage from transmitting input features. More recent work~\cite{pan2025federated} refines GA by stabilizing the loss and deriving update directions that emphasize the forgotten client while minimally conflicting with retained ones, reducing utility degradation.

\mypara{Vertical Federated Unlearning}  
While our work focuses on horizontal FL, vertical FL has also gained attention, where different parties hold complementary feature spaces.  
Recent studies propose efficient and privacy-aware VFL training frameworks~\cite{liu2024vertical, zhang2024s2nerf}, but unlearning in VFL~\cite{wang2024efficient, han2025vertical} is particularly challenging since deleting one user's data requires coordinated removal across multiple parties.  
Existing approaches such as Split-U~\cite{yu2023split} provide early solutions but still incur notable computational or communication cost.

\mypara{Machine Unlearning Privacy Leakage}  
Recent work~\cite{chen2021machine} shows that the unlearning process itself may introduce new privacy risks, as adversaries can potentially infer membership or sensitive attributes of the unlearned data from model updates.  
Such leakage risks naturally extend to the federated setting~\cite{wang2023federated}, where distributed updates and client-server interactions may further amplify the exposure surface.  
Although our work focuses on exact unlearning correctness and efficiency, mitigating privacy leakage during the unlearning process remains an important direction for future federated unlearning research.

\mypara{Privacy-preserving FL}
FL raises privacy concerns~\cite{wei2020federated, liu2024cross, zhu2019deep, liu2020fedsel}, since exchanged updates can leak information about client data through gradient or activation inversion. While our work focuses on FU, we note that the training stage of most FU frameworks inherits the same privacy risks as standard FL. 
Improving training-time privacy (e.g., via differential privacy~\cite{dwork2014algorithmic}) is largely orthogonal to our FU design and represents a promising future work.

\section{Discussion}
\label{appendix:discussion}


\mypara{Efficient Retraining}
Although our primary objective is to maintain uninterrupted service and minimize retraining whenever possible, certain applications may still require partial model recovery after several deletions to preserve performance. This approach is preferable to retraining the entire model from scratch once all data groups or clusters are affected, as otherwise the system may experience performance degradation due to a reduced training dataset size.
FedCIO supports efficient retraining by updating only the clients within the affected cluster. Similarly, \method can accommodate retraining through both online and offline processes. Specifically, \method first identifies the sequence with the least number of affected groups and performs online retraining on those groups, excluding the unlearned data points. This minimizes the training cost while recovering model performance on the remaining dataset. In parallel, other sequences are retrained offline to proactively prepare for potential future unlearning requests.

\mypara{Unlearning Request Distribution}
In this paper, we assume that unlearning requests arrive randomly, meaning the deleted data samples are independent across requests.
However, in certain real-world scenarios, unlearning requests may be related.
For example, in a healthcare FL system involving multiple hospitals, it is plausible that one hospital frequently issues consecutive unlearning requests, while others rarely do so.

Such unlearning patterns can be incorporated into \method's design.
By adjusting the grouping strategy so that data from highly related clients are more likely to fall into the same group, \method can reduce the number of affected groups when deletions occur.
Concentrating related data into fewer groups allows the system to preserve more valid sequences after each unlearning request, thereby improving robustness and reducing degradation of model utility.

\section{EXPERIMENTAL SETUP}\label{appendix:exp_setup}
Dataset-specific hyperparameters are summarized in \autoref{tab:hyperparams}.

\end{document}